\documentclass[12pt,preprint]{aastex}

\def\kms{\ifmmode{\rm km\thinspace s^{-1}}\else km\thinspace s$^{-1}$\fi}
\def\V10{V1061\,Cyg}

\slugcomment{To appear in 01 Apr 2006 issue of ApJ}

\shortauthors{Torres et al.}
\shorttitle{V1061 Cyg}

\begin{document}

\title{The Eclipsing Binary V1061 Cygni: Confronting Stellar Evolution
Models for Active and Inactive Solar-Type Stars}

\author{Guillermo Torres\altaffilmark{1},
	Claud H.\ Sandberg Lacy\altaffilmark{2},
	Laurence A.\ Marschall\altaffilmark{3},
	Holly A.\ Sheets\altaffilmark{4}, and
	Jeff A.\ Mader\altaffilmark{5}
}

\altaffiltext{1}{Harvard-Smithsonian Center for Astrophysics, 60
Garden St., Cambridge, MA 02138, e-mail: gtorres@cfa.harvard.edu}

\altaffiltext{2}{Department of Physics, University of Arkansas,
Fayetteville, AR 72701, e-mail: clacy@uark.edu}

\altaffiltext{3}{Department of Physics, Gettysburg College, 300 North
Washington Street, Gettysburg, PA 17325, e-mail:
marschal@gettysburg.edu}

\altaffiltext{4}{Department of Physics \& Astronomy, Dartmouth
College, Hanover, NH 03755, e-mail: Holly.A.Sheets@Dartmouth.edu}

\altaffiltext{5}{W.\ M.\ Keck Observatory, Kamuela, HI 96743, e-mail:
jmader@keck.hawaii.edu}

\begin{abstract} 

We present spectroscopic and photometric observations of the eclipsing
system \V10, previously thought to be a member of the rare class of
``cool Algols". We show that it is instead a hierarchical triple
system in which the inner eclipsing pair (with $P = 2.35$ days) is
composed of main-sequence stars and is well detached, and the third
star is also visible in the spectrum. We combine the radial velocities
for the three stars, times of eclipse, and intermediate astrometric
data from the HIPPARCOS mission (abscissae residuals) to establish the
elements of the outer orbit, which is eccentric and has a period of
15.8 yr. We determine accurate values for the masses, radii, and
effective temperatures of the binary components: $M_{\rm Aa} = 1.282
\pm 0.015$~M$_{\sun}$, $R_{\rm Aa} = 1.615 \pm 0.017$~R$_{\sun}$, and
$T_{\rm eff}^{\rm Aa} = 6180 \pm 100$~K for the primary (star Aa), and
$M_{\rm Ab} = 0.9315 \pm 0.0068$~M$_{\sun}$, $R_{\rm Ab} = 0.974 \pm
0.020$~R$_{\sun}$, and $T_{\rm eff}^{\rm Ab} = 5300 \pm 150$~K for the
secondary (Ab). The masses and radii have relative errors of only
1--2\%.  Both stars are rotating rapidly ($v \sin i$ values are $36
\pm 2$~\kms\ and $20 \pm 3$~\kms) and have their rotation synchronized
with the orbital motion.  There are signs of activity including strong
X-ray emission and possibly spots.  The mass of the tertiary is
determined to be $M_{\rm B} = 0.925 \pm 0.036$~M$_{\sun}$ and its
effective temperature is $T_{\rm eff}^{\rm B} = 5670 \pm 150$~K. The
system is at a distance of $166.9 \pm 5.6$ pc.  Current stellar
evolution models that use a mixing length parameter $\alpha_{\rm ML}$
appropriate for the Sun agree well with the properties of the primary,
but show a very large discrepancy in the radius of the secondary, in
the sense that the predicted values are $\sim$10\% smaller than
observed (a $\sim$5$\sigma$ effect). In addition, the temperature is
cooler than predicted by some 200~K. These discrepancies are quite
remarkable given that the star is only 7\% less massive than the Sun,
the calibration point of all stellar models. Similar differences have
been seen before for later-type stars, but the source of the problem
has remained unclear. A comparison with the properties of other stars
of similar mass as the secondary in \V10 has allowed us to identify
the chromospheric activity as the likely cause of the effect.
Inactive stars agree very well with the models, while active ones such
as \V10 Ab appear systematically too large and too cool. Theory
provides an understanding of this in terms of the strong magnetic
fields commonly associated with stellar activity, which tend to
inhibit convective heat transport.  The reduced convection explains
why fits to models with a smaller mixing length parameter of
$\alpha_{\rm ML} = 1.0$ seem to give better agreement with the
observations for \V10~Ab.

\end{abstract}

\keywords{binaries: close --- binaries: spectroscopic --- stars:
evolution --- stars: individual (\V10) --- techniques: spectroscopic}

\section{Introduction}
\label{sec:introduction}

Accurately determined properties of stars in detached eclipsing
binaries provide fundamental data for testing models of stellar
structure and stellar evolution \citep[see, e.g.,][]{Andersen:91,
Andersen:97}. For stars less massive than the Sun properties such as
the stellar radius and the effective temperature have occasionally
been found to disagree with model predictions \citep[see,
e.g.,][]{Lacy:77, Popper:97, Clausen:99a, Torres:02, Ribas:03}.
Directed efforts to find additional systems in this regime suitable
for testing theory \citep{Popper:96, Clausen:99b} have produced a few
cases, while other examples have been found serendipitously
\citep[e.g.,][]{Creevey:05, Lopez-Morales:05}. The present binary
system is in the second category, since it was originally thought to
be of a completely different nature.
	
The photometric variability of V1061 Cygni (also known as HD~235444,
HIP~104263, RX\,J2107.3+5202, $\alpha = 21^{\rm h} 07^{\rm m}
20\fs52$, $\delta = +52\arcdeg 02\arcmin 58\farcs4$, J2000, SpT F9, $V
= 9.24$) was discovered photographically by \cite{Strohmeier:59}, and
the object was classified by \cite{Strohmeier:62} as an Algol-type
binary with a period of 2.346656 days.  Other than occasional
measurements of the time of primary eclipse, the system received very
little attention until the spectroscopic work by \cite{Popper:96}, who
observed it as part of his program to search for eclipsing binaries
containing at least one lower main-sequence star (late F to K).  On
the basis of two high-resolution spectra and other information Popper
concluded that \V10 was most likely a semi-detached system of the rare
``cool Algol" class, and dropped it from his program. Unlike the
classical Algols, which are composed of a cool giant or subgiant and
an early-type star, the mass-gainer in the cool Algols is also of late
spectral type \citep[see][]{Popper:92}.  Since less than a dozen of
these systems are known, \V10 was placed on the observing list at the
Harvard-Smithsonian Center for Astrophysics (CfA) in 1998 for
spectroscopic monitoring, and photometric observations began later.
Not only did we discover that it is not a cool Algol \citep[it is well
detached, as reported by][]{Sheets:03}, but we also found that:
\emph{i)} it is triple-lined (and a hierarchical triple); \emph{ii)}
the secondary in the eclipsing pair is less massive than the Sun and
therefore potentially interesting for constraining models of stellar
structure and evolution (Popper's original motivation for observing
it); and \emph{iii)} the mass ratio of the binary is quite different
from unity, which makes it a favorable case for such tests.
Furthermore, the comparison with theory shows a significant
discrepancy in the radius of the secondary, corroborating similar
evidence from other systems and providing some insight into the
problem.

We describe below our observations and complete analysis of this
system, including a discussion of the possible nature of the
deviations from the models for low-mass stars. 
	
\section{Observations and reductions}
\label{sec:observations}

\subsection{Spectroscopy}
\label{sec:spectroscopy}

\begin{figure}
\epsscale{1.0}
\vskip -1.5in 
\plotone{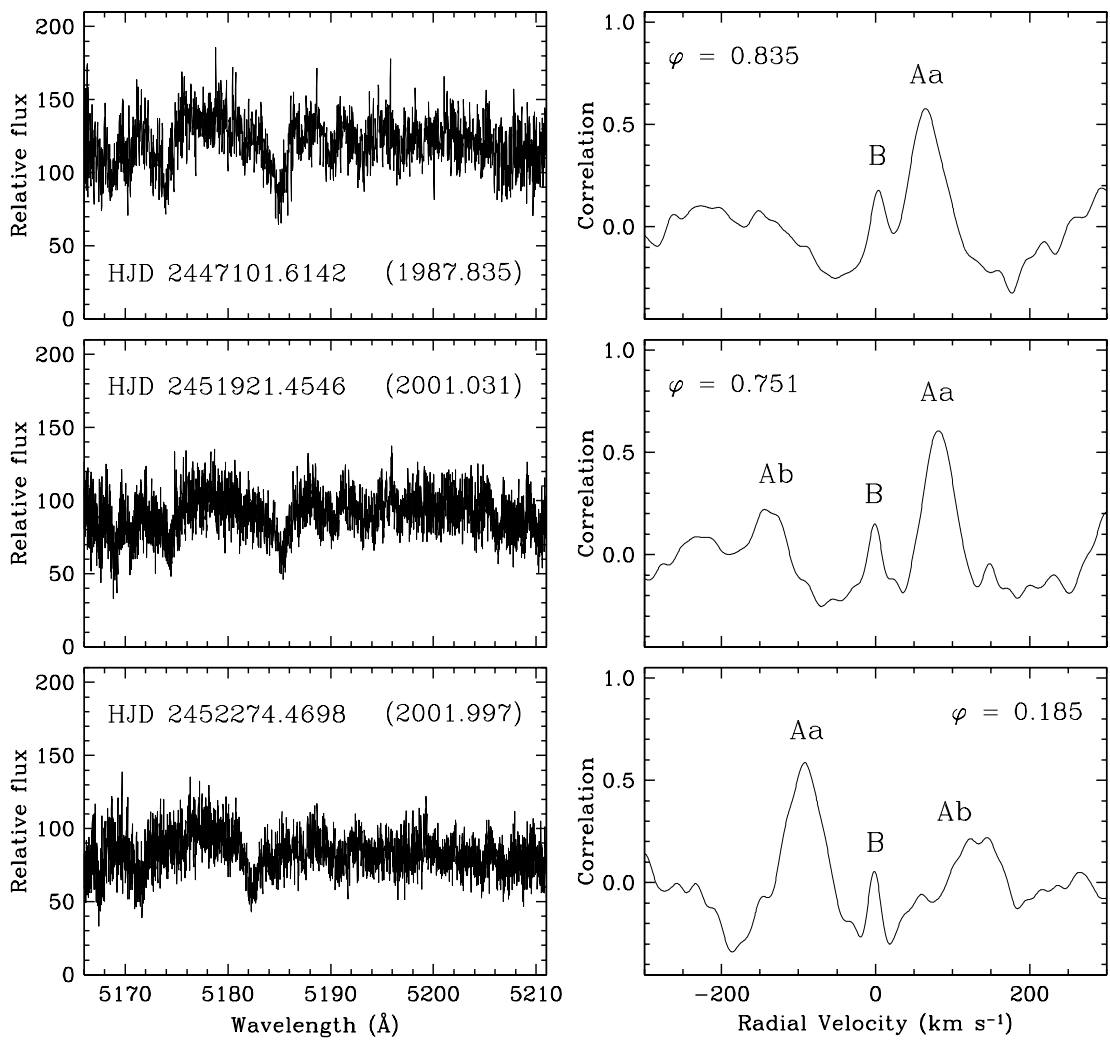}
 \figcaption[]{Sample spectra of \V10 (left panels) and
corresponding cross-correlation functions (right) showing a
broad-lined (Aa) and a sharp-lined (B) component. The Julian date,
year, and orbital phase ($\varphi$) are indicated.  The tertiary star
(Ab) is not obvious in the 1987 spectrum, but is in many of the more
recent observations.\label{fig:spectra}}
 \end{figure}

\V10 was observed at the CfA with an echelle spectrograph on the 1.5-m
Wyeth reflector at the Oak Ridge Observatory (eastern Massachusetts),
and occasionally also with a nearly identical instrument on the 1.5-m
Tillinghast reflector at the F.\ L.\ Whipple Observatory (Arizona).  A
single echelle order was recorded using intensified photon-counting
Reticon detectors, spanning about 45\,\AA\ at a central wavelength of
5187\,\AA, which includes the \ion{Mg}{1}~b triplet. The resolving
power of these instruments is $\lambda/\Delta\lambda \approx
35,\!000$, and the signal-to-noise (S/N) ratios achieved range from 13
to about 40 per resolution element of 8.5~\kms.  A total of 74 spectra
were collected from August 1998 until April 2005. One archival
spectrum from 1987 obtained with the same instrumentation was used as
well.  That observation was taken by J.\ Andersen, possibly at the
request of D.\ Popper (Andersen, priv.\ comm.), and it shows double
lines (one sharp and one broad; see Figure~\ref{fig:spectra}).
\cite{Popper:96} referred to that observation, and remarked that his
own two spectra of the star also showed a sharp-lined and a
broad-lined component. Most of our more recent observations display
similar features. The fact that the sharp-lined star is hardly moving
whereas the other one moves significantly led Popper to infer a mass
ratio of roughly 8/1, and to conclude that \V10 is a cool Algol.

As it turns out, the sharp-lined component is not the secondary in the
eclipsing binary. Closer inspection of the cross-correlation
functions, after we had estimated crude velocities and discovered that
the sharp-lined and broad-lined stars were not moving together,
revealed the presence of a third star (the true secondary) changing
velocity with the same 2.35-day period as the broad-lined
component. This is illustrated in Figure~\ref{fig:spectra}. In the
following we refer to the broad-lined primary of the eclipsing binary
(which is the more massive one of the pair) as star Aa, to the
secondary as star Ab, and to the tertiary as star B.

Radial velocities for the three stars were derived using an extension
of the two-dimens\-ional cross-correlation algorithm TODCOR
\citep{Zucker:94} to three dimensions \citep{Zucker:95}. This
technique uses three templates, one for each star, and is well suited
to our relatively low S/N spectra allowing velocities to be obtained
reliably even when the spectral lines are blended. Templates for the
cross correlations were selected from an extensive library of
calculated spectra based on model atmospheres by R.\ L.\
Kurucz\footnote{Available at {\tt http://cfaku5.cfa.harvard.edu}.},
computed for us by Jon Morse \citep[see
also][]{Nordstrom:94,Latham:02}. These calculated spectra are
available for a wide range of effective temperatures ($T_{\rm eff}$),
projected rotational velocities ($v \sin i$), surface gravities ($\log
g$), and metallicities. Experience has shown that radial velocities
are largely insensitive to the surface gravity and metallicity adopted
for the templates. Consequently, the optimum template for each star
was determined from grids of cross-correlations over broad ranges in
temperature and rotational velocity (since these are the parameters
that affect the radial velocities the most), seeking to maximize the
average correlation weighted by the strength of each exposure. Surface
gravities of $\log g = 4.0$ and $\log g = 4.5$ were adopted for Aa and
Ab, from the light curve analysis described in \S\ref{sec:photometry},
and for star B we adopted $\log g = 4.5$.  Solar metallicity was
assumed throughout, which the discussion in \S\ref{sec:discussion}
shows to be a good assumption in this case.  The procedure of
optimizing the templates leads to best-fit parameters for the stars as
follows:

 \begin{tabular}{llllllcccr}
&&&&& Aa: & $T_{\rm eff} = 6160 \pm 100$~K & , & $v \sin i =$ & $36 \pm 2~\kms$ \\
&&&&& Ab: & $T_{\rm eff} = 5400 \pm 200$~K & , & $v \sin i =$ & $20 \pm 3~\kms$ \\
&&&&& B:  & $T_{\rm eff} = 5670 \pm 150$~K & , & $v \sin i =$ & $ 2 \pm 3~\kms$ \\
 \end{tabular}

The stability of the zero-point of our velocity system was monitored
by means of exposures of the dusk and dawn sky, and small run-to-run
corrections were applied in the manner described by \cite{Latham:92}.
In addition to the radial velocities, we derived the light
contributions of each star following \cite{Zucker:95}. These are
$\ell_{\rm Aa} = 0.75 \pm 0.01$, $\ell_{\rm Ab} = 0.09 \pm 0.01$, and
$\ell_{\rm B} = 0.16 \pm 0.01$, expressed as fractions of the total
light, and they correspond to the mean wavelength of our observations
(5187\,\AA). 

Due to the narrow wavelength coverage of the CfA spectra there is
always the possibility of systematic errors in the velocities,
resulting from lines of the stars moving in and out of the window with
orbital phase \citep{Latham:96}.  Occasionally these errors are
significant, and experience has shown that this must be checked on a
case-by-case basis \citep[see, e.g.,][]{Torres:97, Torres:00}. For
this we performed numerical simulations in which we generated
artificial composite spectra by adding together synthetic spectra for
the three components, with Doppler shifts appropriate for each actual
time of observation, computed from a preliminary orbital solution.
The light fractions adopted were those derived above.  We then
processed these simulated spectra with the three-dimensional version
of TODCOR in the same manner as the real spectra, and compared the
input and output velocities. The differences were typically small for
stars Aa and B (under 1~\kms), and were up to about 3~\kms\ for star
Ab. We applied these differences as corrections to the raw velocities,
although they made little difference in the results other than
slightly decreasing the residuals from the orbital fits. The final
velocities including these corrections are given in
Table~\ref{tab:rvs}.  Similar corrections were derived for the light
fractions, and are already accounted for in the values reported above.
As a further test we changed the template parameters for the two
fainter stars, which are more uncertain, and recomputed the radial
velocities and corrections. The resulting velocities were hardly
different, giving us confidence in the robustness of the procedures.

\begin{figure} 
\epsscale{0.9} 
\plotone{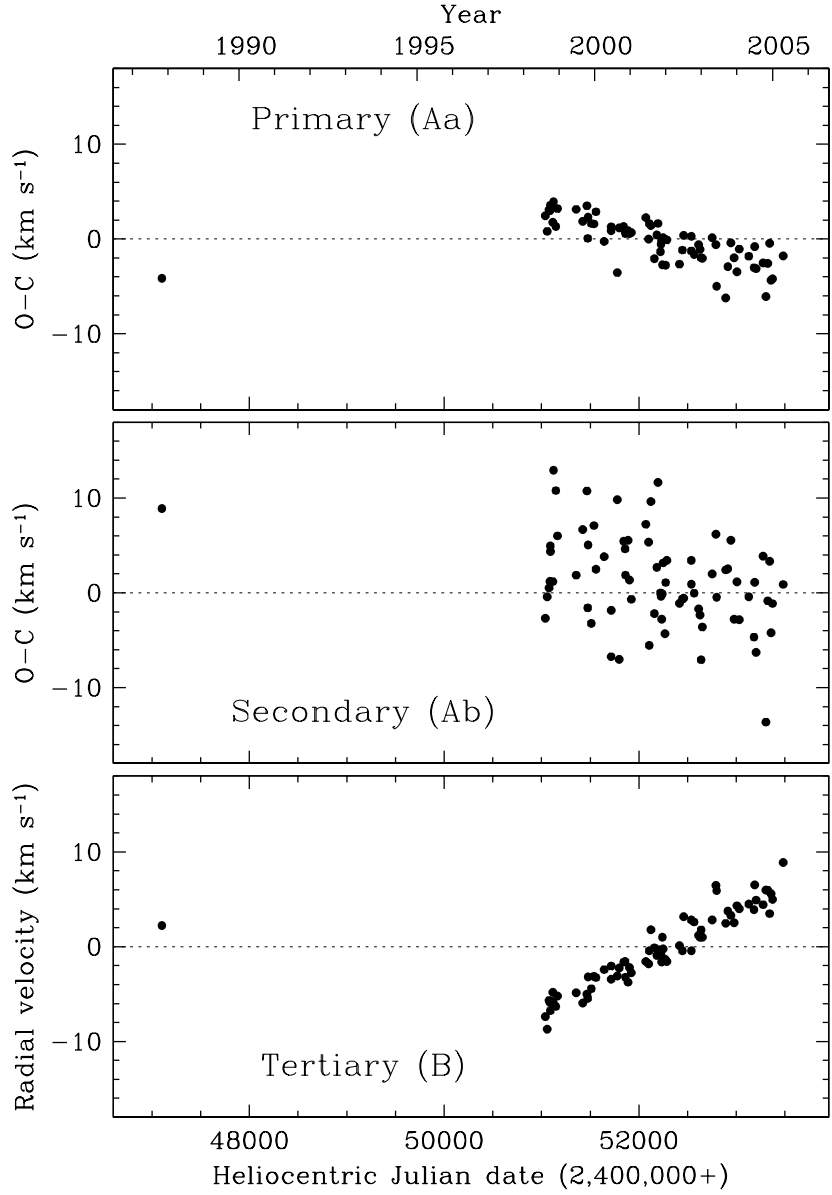}
\vskip 0.5in
 \figcaption[]{Radial velocity residuals for the primary
and secondary of \V10 from a preliminary double-lined orbital
solution, and radial velocities for the tertiary. The trends indicate
the system is a hierarchical triple.\label{fig:rvresiduals}}
 \end{figure}

A double-lined spectroscopic orbital solution for the Aa+Ab system
(see column~2 of Table~\ref{tab:allelements} below) produced a
reasonably good fit with rms residuals of 2.4 and 5.1~\kms, at about
the level expected based on the broad lines and faint secondary.
However, the residuals showed an obvious downward trend, less obvious
but still present also in the secondary. The velocities for star B
show a drift in the opposite direction (Figure~\ref{fig:rvresiduals})
with a slope about twice as steep, or $\sim$5~\kms\ per 1000 days. The
system is thus a hierarchical triple, with a period for the outer
orbit that could be very long, judging from the data available. The
spectroscopic material alone does not sufficiently constrain the wide
orbit to properly account for the change in the center-of-mass
velocity of the eclipsing pair, which could introduce biases in the
derived masses of components Aa and Ab.  However, as we show in the
next section, the additional information provided by the measurement
of times of eclipse and, to some extent, the astrometric data
\citep{ESA:97} from the HIPPARCOS mission described later are very
helpful in constraining the elements of the outer orbit.

\subsection{Times of eclipse}
\label{sec:times}

\begin{figure} 
\vskip -1in
\epsscale{1.0} 
\plotone{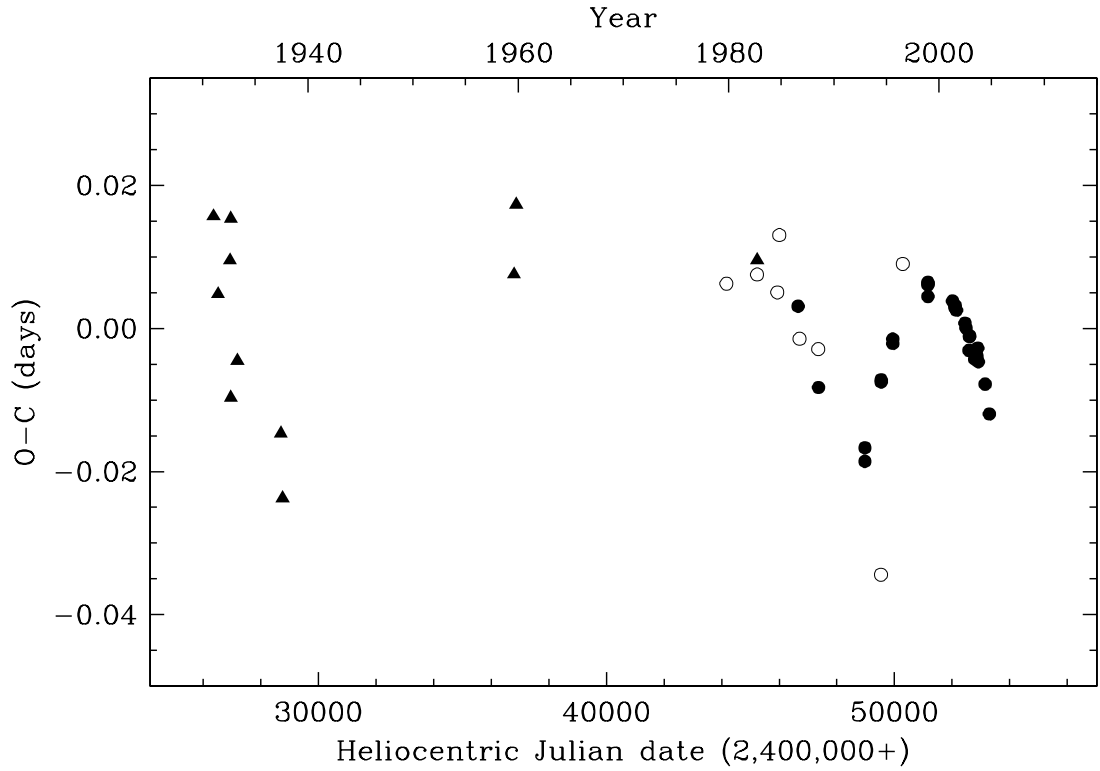}
\vskip -1in
 \figcaption[]{$O-C$ diagram of all available times of
eclipse for \V10, from a preliminary linear ephemeris derived from a
fit to the photoelectric (filled circles), visual (open circles) and
photographic measurements (triangles) for both the primary and
secondary minima. Oscillations suggesting a third body are obvious,
particularly in the recent data.\label{fig:origomc}}
 \end{figure}

Measurements of the times of minimum light for \V10 spanning more than
seven decades have been made by photographic, visual, and
photoelectric/CCD techniques. Table~\ref{tab:timings} collects all
available estimates, including some that have not appeared previously
in print and were kindly communicated to us by J.\ M.\ Kreiner
\citep[see][]{Kreiner:00}. A total of 43 timings for the deeper
primary minimum and 6 for the secondary are listed.  Many of the more
recent measurements between 2001 and 2004 are based on our own
photometric observations with a robotic telescope described in more
detail below. The $O-C$ residuals from a linear ephemeris are shown in
Figure~\ref{fig:origomc}, and exhibit a pattern of variations caused
by the presence of the distant third component (light travel effect).
This is most obvious in the more accurate and numerous measurements of
the last 25 years, during which more than a full cycle has been
covered.  The period of the outer orbit is therefore fairly well
constrained by these data, as is the amplitude of the variation, which
is directly related to the velocity amplitude of the binary around the
center-of-mass of the triple. These measurements provide information
complementary to that from the radial velocities, and we incorporate
them into the analysis in \S\ref{sec:orbit}.

\subsection{Photometry}
\label{sec:photometry}

Absolute photometry of \V10 on the $UBV$ system was obtained by
\cite{Lacy:92}. The value of the reddening-free parameter $Q = (U-B) -
0.72 (B-V)$ was calculated and compared with the standard star values
of \cite{Johnson:53}. The value corresponded to that of a spectral
type G0 star with an intrinsic color of $(B-V)_0 =
0.60^{+0.00}_{-0.10}$, with no significant interstellar reddening. The
uncertainty here comes mainly from the calibration. The observed mean
color index corresponds to a spectral type F9 star with a temperature
of 6040~K, according to the calibration of \cite{Popper:80}. After
some iterations with the photometric light curve fitting algorithm
described below, it was found that the visual surface brightness ratio
$J_{\rm Ab}/J_{\rm Aa}$ is about 0.43, which constrains the difference
in the visual flux parameter $F^{\prime}_V$ and hence the temperature
difference through the calibration in Table~1 by \cite{Popper:80}.
This procedure leads to photometric temperature estimates of $6200 \pm
100$~K for the primary and $5280 \pm 100$~K for the secondary
component. These temperatures were used to estimate the limb-darkening
and gravity-brightening parameters to be used in the photometric
modeling. They are in excellent agreement with those of the
spectroscopic analysis above.
	
Relative photometric observations of \V10 were obtained at two
different facilities. Measurements in the $V$ band were made with the
0.26-m URSA robotic telescope at Kimpel Observatory on the campus of
the University of Arkansas at Fayetteville. A total of 6914
observations were collected between March 2001 and October 2004, which
are given in Table~\ref{tab:ursa}. For details on the observation and
reduction methods the reader is referred to the description by
\cite{LacyCS:04}. The precision of an individual measurement is
estimated to be 0.010 mag. Observations were also made with the 0.4-m
f/11 Ealing Cassegrain reflector at Gettysburg College Observatory, in
Gettysburg (Pennsylvania). The detector was a Photometrics (Roper
Scientific) CH-350 thermoelectrically-cooled CCD camera with a
back-illuminated SITe 003B $1024 \times 1024$ chip and standard
Bessell $BV\!RI$ filters.  Tables~\ref{tab:gettyb}--\ref{tab:gettyi}
list the observations obtained between September 2002 and September
2003, which number 732, 738, 739, and 740 in the $B$, $V$, $R$, and
$I$ passbands, respectively. The estimated precision of these
differential measurements is 0.011 mag in $B$, 0.008 mag in $V$, 0.009
mag in $R$, and 0.012 mag in $I$.  The comparison stars used for these
two data sets are given in Table~\ref{tab:comparisons}.  The URSA
photometry was referenced to the magnitude corresponding to the sum of
the intensities of all four comparison stars in order to improve the
measurement accuracy.

Because the eclipse timings for \V10 described earlier show changes
due to motion about the center of mass of the triple system, light
elements used in analyzing the photometric data were based on
contemporary observations made between March 2001 and October 2004
with the URSA telescope. As seen in Figure~\ref{fig:origomc} the
curvature of the $O-C$ residuals is relatively small in this interval,
and a linear ephemeris is sufficient for our purposes. A fit to 12
primary minima (see Table~\ref{tab:timings}) yielded
\begin{displaymath} {\rm Min~I~(HJD)} = 2,\!452,\!015.90562(7) +
2.34663383(33) \cdot E~, \end{displaymath} where $E$ is the number of
cycles elapsed from the epoch of reference, and the values in
parentheses represent the uncertainties of the elements in units of
the last digit.  Photometric data were then phased according to these
elements.

The URSA and Gettysburg data sets were analyzed with the
Nelson-Davis-Etzel model (EBOP code) as described by \cite{Etzel:81}
and \cite{Popper:81}. Although other models may be more sophisticated,
this program is perfectly adequate for well-detached systems such as
\V10. The main adjustable parameters in the NDE model are the relative
surface brightness ($J \equiv J_{\rm Ab}/J_{\rm Aa}$) of the secondary
of the eclipsing binary (star Ab) in units of that of the primary
(Aa), the relative radius of the primary ($r_{\rm Aa}$) in units of
the separation, the ratio of radii ($k \equiv r_{\rm Ab}/r_{\rm Aa}$),
the inclination of the orbit of the binary ($i_{\rm A}$), the
limb-darkening coefficients ($x_{\rm Aa}$ and $x_{\rm Ab}$), the
eccentricity parameters $e \cos\omega$ and $e \sin\omega$, and the
amount of third light ($\ell_{\rm B}$, expressed as a fraction of the
total light), which is significant in our case.  The luminosity due to
the reflection effect was computed from bolometric theory. The mass
ratio ($q = 0.7266$) was adopted from the spectroscopic results in
\S\ref{sec:orbit}.  Tests showed the orbital eccentricity was probably
not significantly different from zero, consistent with the results
described later, so $e$ was fixed at zero for the photometric
modeling. For the URSA data the EBOP algorithm converged with all
variables free, except for the gravity-brightening coefficients $y$,
which were set from theory \citep{Alencar:97, Claret:98}.  For the
Gettysburg data sets the limb-darkening coefficients $x$ had to be
fixed at theoretical values \citep{Al-Naimiy:78, Wade:85,
Diaz-Cordoves:95}, as were the gravity-brightening coefficients.
Additionally, for the Gettysburg data the values of third light in the
spectral bands $B$, $R$, and $I$ had to be estimated from the
temperature of the third star based on the spectroscopic analysis in
\S\ref{sec:spectroscopy}. The third light value in $V$ was adopted
from the URSA data solution ($\ell_{\rm B} = 0.144$).  For these
reasons, the final mean geometric parameters based on the Gettysburg
data have been assigned uncertainties two times the internal
estimates, consistent with our experience with EBOP in these
situations.

Initial light curve solutions with the URSA data showed that
observations during the night of JD $2,\!452,\!054$ (24 May 2001),
which followed a secondary eclipse, were significantly fainter than
the fitted orbit by $\sim$0.025 mag. No other observations at this
phase (0.57) were collected until the 2003--2004
seasons. Additionally, data on JD $2,\!453,\!218$ and JD
$2,\!453,\!232$ (2004 season) obtained during and after secondary
eclipse were seen to be brighter than the fitted curve by $\sim$0.01
mag. Measurements taken 12 days earlier at a similar phase appear to
be normal, as are those taken two months later.  Possible explanations
for these deviations include some unknown calibration problem in the
data reductions, or transitory spots on one or both stars that change
on timescales of a month or two. The latter would not be unusual given
the strong X-ray emission from the system \citep{Voges:99}, indicative
of chromospheric activity (see \S\ref{sec:activity}). If due to spots,
the extent of the active regions appears to be quite large (up to
40\arcdeg--50\arcdeg\ in longitude), judging from the phase interval
affected. Tests showed that removal of these URSA observations had an
insignificant effect on the results other than decreasing the
scatter. We have thus chosen to exclude the data on those three nights
from the final fitting process.  Further support for intrinsic
variations is given by similar deviations seen in the Gettysburg data
at other phases, particularly phase $\sim$0.45, as shown below in
Figure~\ref{fig:getty}. The angular extent of this particular feature
is again $\sim$40\arcdeg. A hint of a dimming at about the same phase
may be present also in the URSA observations during the same season,
although the data sampling is not optimal for confirming this.

\begin{figure} 
\epsscale{1.0} 
\vskip -1.5in
\plotone{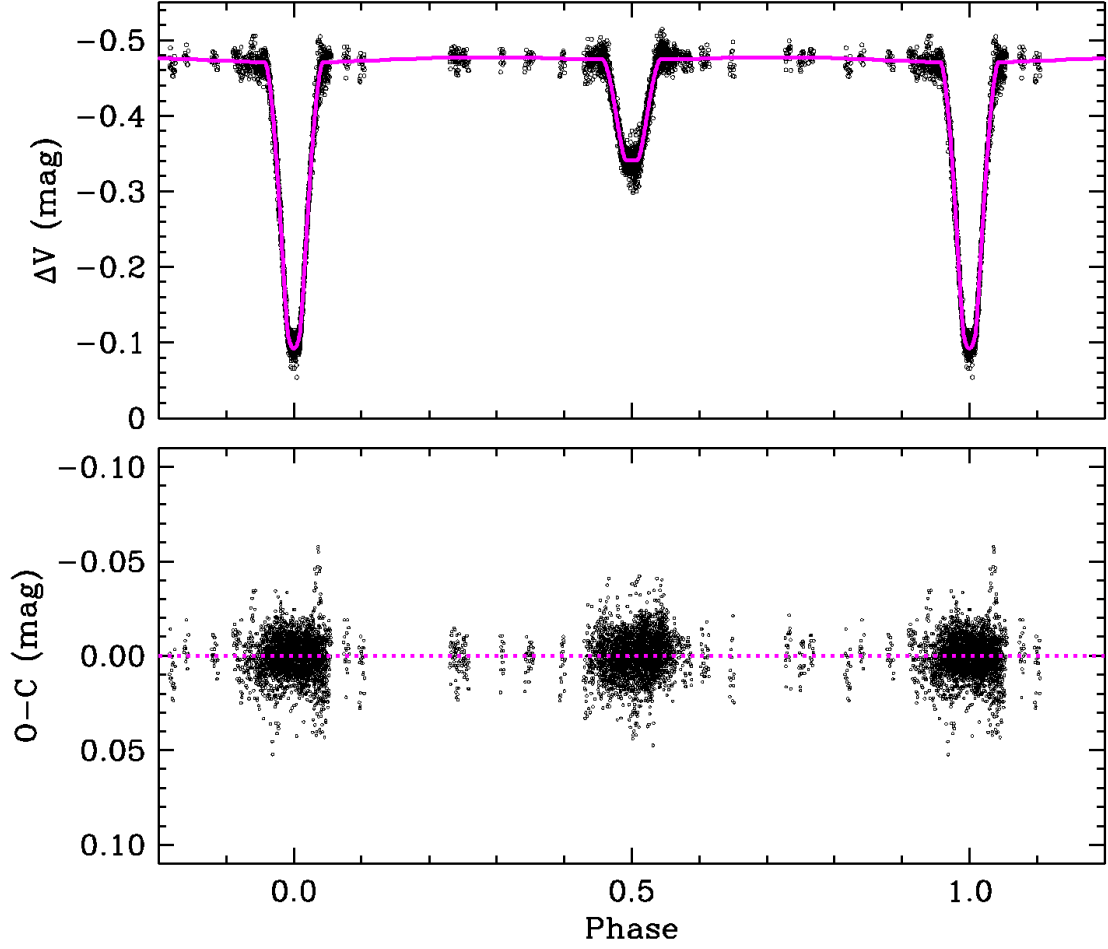}
\vskip -0.5in 
 \figcaption[]{URSA $V$-band photometry for \V10 along
 with our fitted light curve, and residuals from that
 fit.\label{fig:ursaall}}
 \end{figure}

\begin{figure} 
\epsscale{0.9}
\vskip -1.5in
\plotone{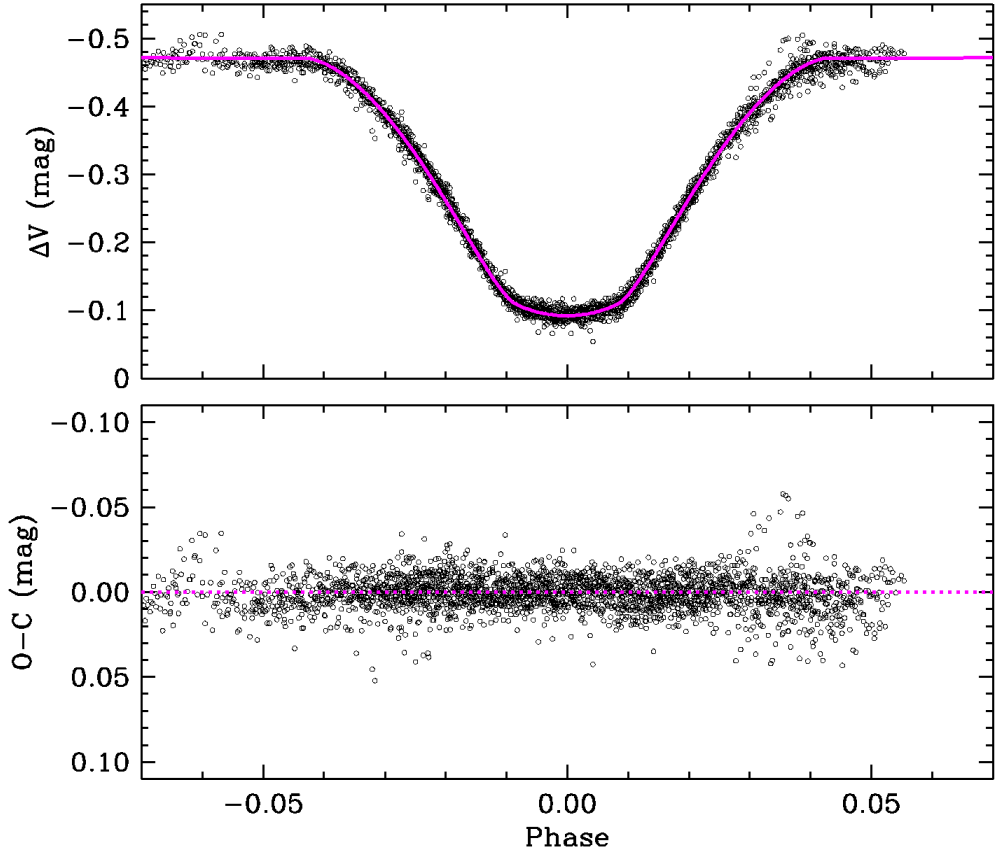}
\vskip -0.5in
 \figcaption[]{URSA $V$-band photometry for \V10 around
the primary minimum, along with our model light curve.  Residuals from
the fit are shown in the bottom panel.\label{fig:ursaprim}}
 \end{figure}

\begin{figure} 
\epsscale{0.9}
\vskip -1.5in
\plotone{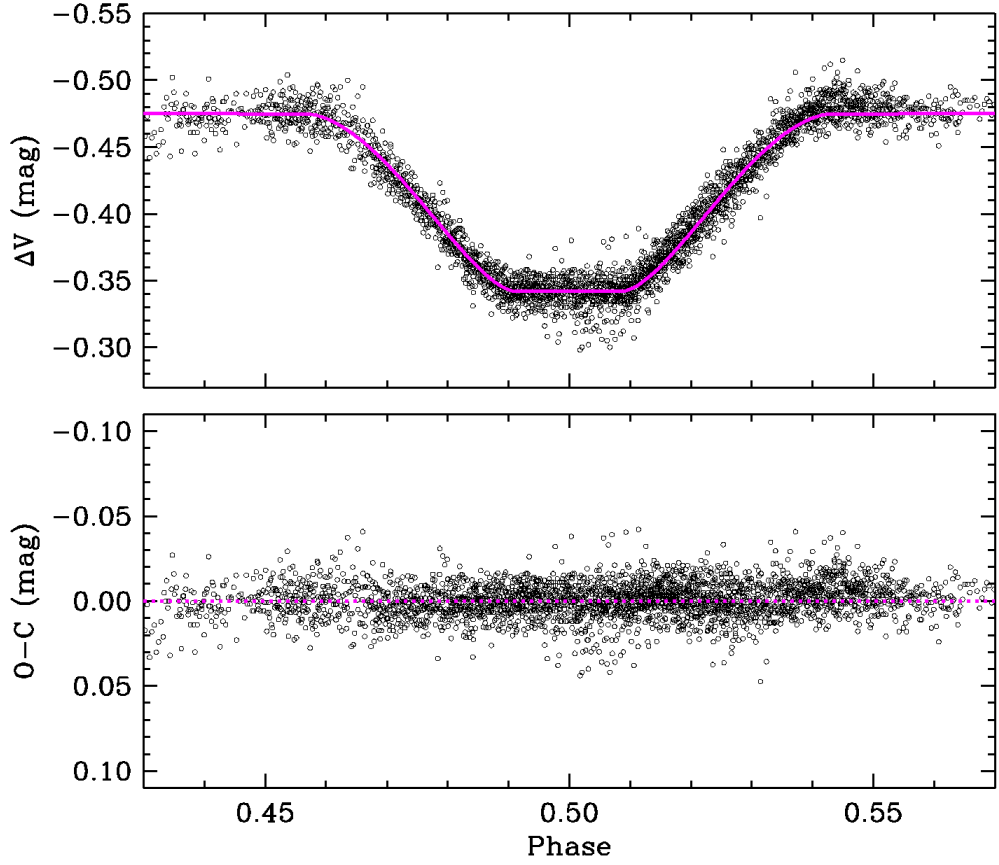}
\vskip -0.5in
 \figcaption[]{URSA $V$-band photometry for \V10 around
the secondary minimum, along with our model light curve. Residuals
from the fit are shown in the bottom panel. Note that the vertical
scale in the top panel is different from
Figure~\ref{fig:ursaprim}.\label{fig:ursasec}}
 \end{figure}

The results of our photometric analysis are given in
Table~\ref{tab:lightelements}, separately for each of the Gettysburg
passbands and for URSA $V$. The mean of the Gettysburg results are
also listed, with uncertainties that account for the scatter between
the $BV\!RI$ results.  There is excellent agreement in the geometric
elements between the two data sets. The light curve and residuals for
the URSA measurements are shown in Figure~\ref{fig:ursaall}, and
enlargements of the regions around the minima are shown in
Figure~\ref{fig:ursaprim} and Figure~\ref{fig:ursasec}. The Gettysburg
light curves and residuals are shown in Figure~\ref{fig:getty}.  The
binary is well detached: the stars occupy only $\sim$7\% and $\sim$2\%
by volume of their critical equipotential surfaces.  Both components
are nearly spherical in shape \citep[oblateness $< 0.005$, as defined
by][]{Etzel:81}. The secondary eclipse is total, and the fraction of
the primary light blocked at phase 0.0 is 40\%.

A common problem in solving the light curves of some eclipsing systems
is the indeterminacy in the ratio of the radii, particularly when the
components are similar and the eclipses partial. As a test, we fitted
the URSA light curves with fixed values of $k$ over a wide range
around the best-fit value of 0.603.  Solutions converged only between
0.55 and 0.63. Some parameters such as $\ell_{\rm B}$, and to a lesser
extent the relative brightness of the secondary ($\ell_{\rm Ab}$),
showed a very strong sensitivity to $k$. Only fits in the vicinity of
$k = 0.603$ (which minimizes the sum of residuals squared) resulted in
values of $\ell_{\rm B}$ and $\ell_{\rm Ab}$ consistent with the
spectroscopic estimates from \S\ref{sec:spectroscopy} (after a small
correction from the mean wavelength of 5187\,\AA\ to the $V$ band). We
take this as an indication that the adopted fits are realistic.
	
\subsection{Astrometry}
\label{sec:astrometry}

\V10 was observed by the HIPPARCOS satellite under the designation
HIP~104263, from December 1989 through February 1993. The measured
trigonometric parallax is $\pi_{\rm HIP} = 6.25 \pm 1.06$~mas.  The
photometric variability of the star with the known 2.35-day period was
clearly detected in the more than 120 brightness measurements made
over the duration of the mission.  Additionally, even though the wide
orbit in \V10 is much too small to be spatially resolved by HIPPARCOS,
the astrometric solution revealed perturbations in the motion across
the sky that led to its classification as a ``Variability-Induced
Mover" (VIM). These perturbations arise from the fact that the center
of light of the triple system shifts during the eclipses due to the
change in the relative brightness between the binary and the third
star. The shift is approximately 25 milli-arc seconds, which the
satellite was able to detect. As a first-order approximation the
HIPPARCOS team assumed the angular separation between the binary and
the third star was constant, so that the motion of the photocenter is
along a straight line connecting them.  With this they were able to
infer a lower limit to the separation of approximately 70 mas, and a
position angle (North through East) for the third star relative to the
binary of $319\arcdeg \pm 14\arcdeg$.  This linear and periodic motion
of the photocenter was accounted for in deriving the position, proper
motion, and parallax of \V10, as reported in the catalog.

In reality the third star and the binary are in orbit around each
other, and from Figure~\ref{fig:origomc} the period appears to be of
the order of 15 years. The 3.2-yr interval of the HIPPARCOS
observations represents a non-negligible fraction of a cycle, and
depending on the orientation of the orbit this may have an impact on
the derived parallax and proper motion, as well as the inferred motion
of the photocenter. In fact, turning the problem around, the HIPPARCOS
observations can potentially provide valuable constraints on the outer
orbit. Thus we incorporate these observations (in the form of the
intermediate astrometric data, or ``abscissae residuals") into the
orbital solution described next. 
	
\section{Orbital solution}
\label{sec:orbit}

Due to the much larger orbit of the third star compared to the
eclipsing binary in \V10, to first order we assume here that the
hierarchical triple system may be separated into an inner orbit and an
outer orbit, the latter treated as a ``binary" composed of the third
star (B) and the center of mass of the eclipsing pair (A).  Orbital
elements that refer to the outer orbit are indicated below with the
subindex ``AB", and those pertaining to the inner orbit are
distinguished with a subindex ``A".

In addition to providing the spectroscopic elements of the inner
orbit, in principle the radial velocity measurements yield also the
elements of the outer orbit, except that in our case the coverage is
insufficient for that purpose (see Figure~\ref{fig:rvresiduals}), and
the period of the outer orbit ($P_{\rm AB}$) is undetermined.
Additionally, the semi-amplitudes of the velocity variation in the
wide orbit ($K_{\rm A}$ and $K_{\rm B}$) are poorly constrained
because the observations do not cover the velocity extrema. The times
of eclipse help in two important ways: they constrain the period
($\sim$15 years from a preliminary analysis), and they constrain the
amplitude of the third body effect ($K_{\rm O-C}$), which is directly
related to $K_{\rm A}$ through $K_{\rm O-C} = K_{\rm A} \sqrt{1-e_{\rm
AB}^2}/2\pi c$, where $e_{\rm AB}$ is the eccentricity of the outer
orbit and $c$ is the speed of light. The fact that the inner binary is
eclipsing allows the masses of the three stars to be determined. For
this we incorporate the angle $i_{\rm A}$ from \S\ref{sec:photometry}
into the solution, along with its uncertainty.

The use of the HIPPARCOS data introduces several other orbital
elements into the problem that are constrained by the astrometry,
including the inclination angle of the wide orbit ($i_{\rm AB}$) and
the position angle of the ascending node ($\Omega_{\rm AB}$, referred
to the equinox of J2000).  Additionally we must solve for corrections
to the catalog values of the position of the barycenter
($\Delta\alpha^*$, $\Delta\delta$) at the mean catalog epoch of
1991.25, corrections to the proper motions ($\Delta\mu_{\alpha}^*$,
$\Delta\mu_{\delta}$), and to the trigonometric parallax
($\Delta\pi$)\footnote{Following the practice in the HIPPARCOS catalog
we define $\Delta\alpha^* \equiv \Delta\alpha \cos\delta$ and
$\Delta\mu_{\alpha}^* \equiv \Delta\mu_{\alpha} \cos\delta$.}.
Because the inner binary is eclipsing there is redundancy in that the
angle $i_{\rm AB}$ can also be derived purely from the spectroscopic
elements and the known value of $i_{\rm A}$. We may thus eliminate one
parameter, and for practical reasons we have chosen to eliminate
$K_{\rm B}$ and to retain $i_{\rm AB}$ explicitly as an adjustable
variable.

Altogether there are 19 unknowns in the least-squares problem:
\{$P_{\rm A}$, $K_{\rm Aa}$, $K_{\rm Ab}$, $e_{\rm A}$, $\omega_{\rm
Aa}$, $T_{\rm A}$\} for the inner orbit, \{$P_{\rm AB}$, $K_{\rm A}$,
$i_{\rm AB}$, $e_{\rm AB}$, $\omega_{\rm A}$, $\Omega_{\rm AB}$,
$T_{\rm AB}$\} for the outer orbit, the center of mass velocity
$\gamma$, and the HIPPARCOS-related elements \{$\Delta\alpha^*$,
$\Delta\delta$, $\Delta\mu_{\alpha}^*$, $\Delta\mu_{\delta}$,
$\Delta\pi$\}. The longitudes of periastron $\omega$ refer to the star
or system indicated; $T_{\rm A}$ is the time of primary eclipse in the
inner orbit, and $T_{\rm AB}$ represents the time of periastron
passage in the outer orbit.
	
\begin{figure} 
\epsscale{0.9}
\plotone{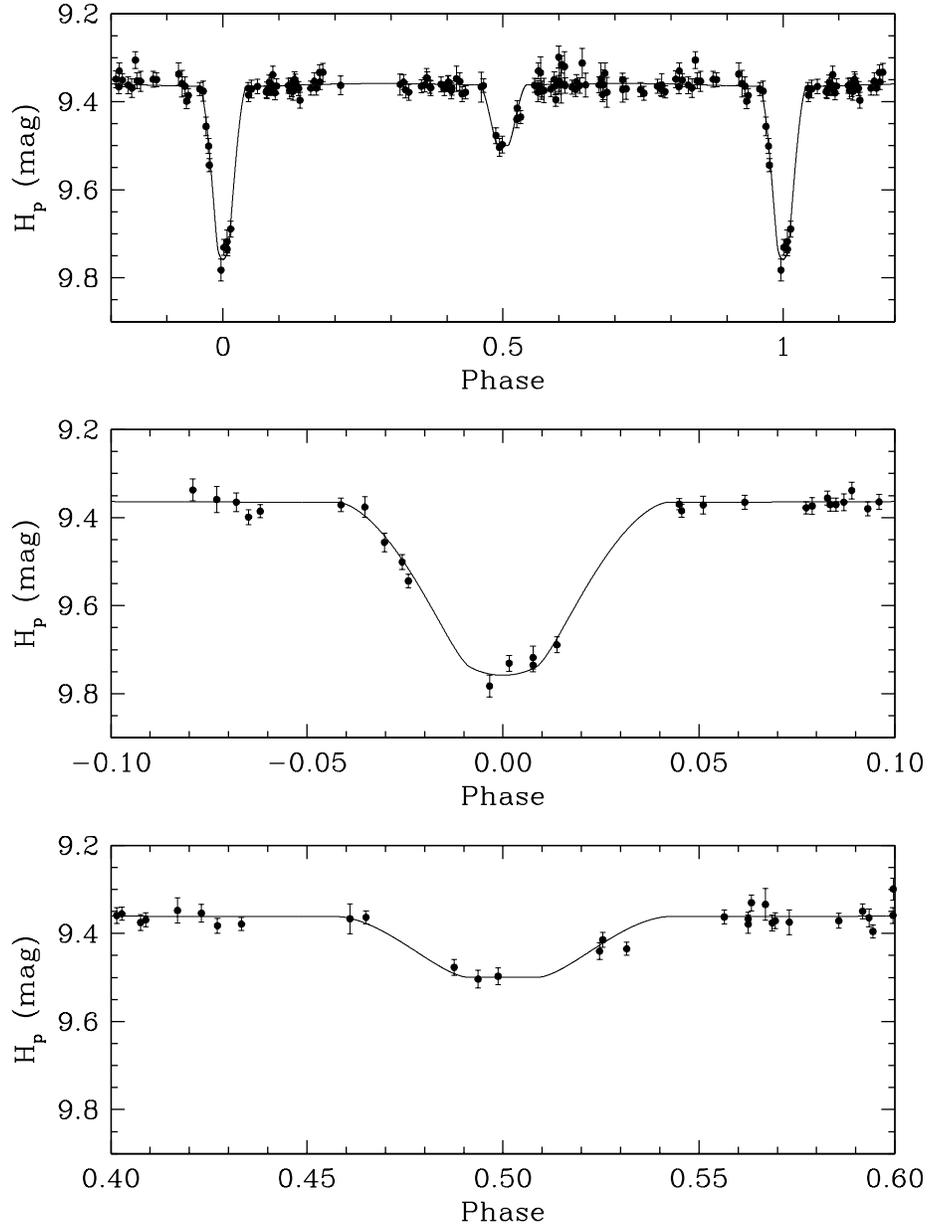}
\vskip 0.5in
 \figcaption[]{HIPPARCOS photometric observations for
\V10 along with our light curve fit using EBOP. The bottom panels show
enlargements of the regions around the minima. This fit was used to
derive the fractional luminosity of star B in the $H_p$ passband (see
text).\label{fig:hiplightcurve}}
 \end{figure}

The amplitude of the motion of the photocenter in angular units,
$\alpha''_{\rm ph}$, is a function of the fractional mass and
luminosity of the third star relative to the system as a whole.  It is
given by $\alpha''_{\rm ph} = a''_{\rm AB} (B-\beta)$, in which the
mass fraction is $B \equiv M_{\rm B}/(M_{\rm Aa} + M_{\rm Ab} + M_{\rm
B})$ and the luminosity fraction is $\beta \equiv L_{\rm B}/(L_{\rm
Aa} + L_{\rm Ab} + L_{\rm B})$. With the definition of third light
used in EBOP, $\beta \equiv \ell_{\rm B}$.  The symbol $a''_{\rm AB}$
represents the relative semimajor axis of the wide orbit expressed in
angular measure. The notation $a''$ is used here to distinguish from
the semimajor axis in linear units, $a$. The mass fraction $B$ can be
expressed in terms of other elements, and $a''_{\rm AB}$ follows from
knowledge of $P_{\rm AB}$, the masses, the parallax, and the
application of Kepler's Third Law.  The luminosity fraction $\ell_{\rm
B}$ needs to be specified in the HIPPARCOS passband ($H_p$), which is
slightly different from Johnson $V$. To derive it we have made use of
the HIPPARCOS epoch photometry for \V10, and solved the light curve
with EBOP.  Since the measurements available are fewer in number
compared to our other light curves, we fixed the geometric elements to
the average of the URSA and Gettysburg results from
Table~\ref{tab:lightelements}, and solved only for the
passband-dependent elements $J$ and $\ell_{\rm B}$.  Limb-darkening
coefficients have little effect and were fixed to their values for the
$V$ band.  The fit to the HIPPARCOS photometry is shown in
Figure~\ref{fig:hiplightcurve}.  The results we obtained from this
adjustment are $\ell_{\rm B} = 0.12 \pm 0.02$ and $J = 0.44 \pm 0.04$
($H_p$).  With this value of $\ell_{\rm B}$ the photocentric semimajor
axis $\alpha''_{\rm ph}$ is completely determined at each iteration of
the least-squares problem and is not an additional free parameter in
the orbital solution.  The formalism for incorporating the abscissae
residuals from HIPPARCOS into the fit follows closely that described
by \cite{vanLeeuwen:98} and \cite{Pourbaix:00}, including the
correlations between measurements from the two independent data
reduction consortia \citep{ESA:97}. 

While the HIPPARCOS observations provide information that is
complementary to that from other available measurements, the
astrometric constraint on some of the orbital elements is not
particularly strong due to the limited precision of those
measurements. Typical uncertainties for a single abscissa residual
range from 2 to 4 mas, which is only a few times smaller than the
motions we are trying to model.  Additional constraints may be placed
on the parallax by using the physical properties of stars Aa and Ab
derived in \S\ref{sec:dimensions}, which allow the absolute luminosity
of each component to be determined very accurately from their
effective temperature and radius. This is the same principle that has
been used to great advantage for applications such as establishing the
distance of external galaxies from eclipsing binaries
\citep[e.g.,][]{Guinan:98, Fitzpatrick:03}. From the radii and
temperatures in \S\ref{sec:dimensions}, along with the total apparent
magnitude of \V10 and the fractional luminosities of the components,
we obtain parallax estimates of $5.90 \pm 0.26$~mas (Aa) and $5.72 \pm
0.48$~mas (Ab), which are consistent with each other. We incorporate
these estimates into the global solution as ``measurements", along
with their uncertainties. 

The delay or advance in the times of eclipse caused by the third star
is accounted for as described by \cite{Irwin:52}. The best-fit time of
primary eclipse ($T_{\rm A}$) given below refers to the center of the
elliptical orbit of the eclipsing pair around the common center of
mass of the triple, as opposed to the barycenter
\citep[see][]{Irwin:52}. For consistency, corrections for light travel
time in the inner binary have also been applied to the dates of the
radial velocity observations at each iteration, which has only a
minimal effect on the spectroscopic elements. These corrections range
from $-0.0145$ days to $+0.0043$ days (a change of 27 minutes over the
6.7-yr timespan of the velocities), and are listed in
Table~\ref{tab:rvs}. 

Standard non-linear least-squares techniques were used to solve
simultaneously for the 19 unknowns that minimize the overall $\chi^2$
of the observations \citep{Press:92}. Weights were assigned to the
measurements according to their individual errors. Most of the older
times of minimum have no published errors, so those were determined by
iterations in order to achieve a reduced $\chi^2$ near unity for each
type of observation (visual, photographic, photoelectric). The adopted
errors are 0.021 days for photographic data, 0.011 days for the visual
timings, and 0.0015 days for photoelectric/CCD minima. The published
errors of the more recent CCD timings were found to require
adjustments to achieve reduced $\chi^2$ values of unity.  The scale
factors used are 0.86 and 3.62 for the primary and secondary minima,
respectively. The larger factor for the secondary is consistent with
the smaller depth of that eclipse. The published uncertainties of the
HIPPARCOS observations were found to require a minor adjustment by a
factor of 0.85, and the optimal errors for the radial velocity
measurements were determined to be 1.51~\kms, 5.40~\kms, and
1.34~\kms\ for stars Aa, Ab, and B, respectively. A total of 332
observations were used: 75 radial velocities for each star, 43 times
of primary eclipse, 6 times of secondary eclipse, 56 HIPPARCOS
observations, and 2 parallax measurements (see above). 

The results of this global fit are given in column~3 of
Table~\ref{tab:allelements}, including derived quantities such as the
masses of the three stars.  The uncertainties listed for all derived
quantities take full account of correlations between different
elements (off-diagonal elements of the covariance matrix).  Initial
solutions allowing for eccentricity in the inner orbit yielded a value
not significantly different from zero, so a circular orbit was
adopted.  The outer orbit is moderately eccentric ($e = 0.469$), and
its period is determined to be 15.8 yr, with an error of only about
1.3\%.

\begin{figure} 
\epsscale{0.9}
\vskip -1in
\plotone{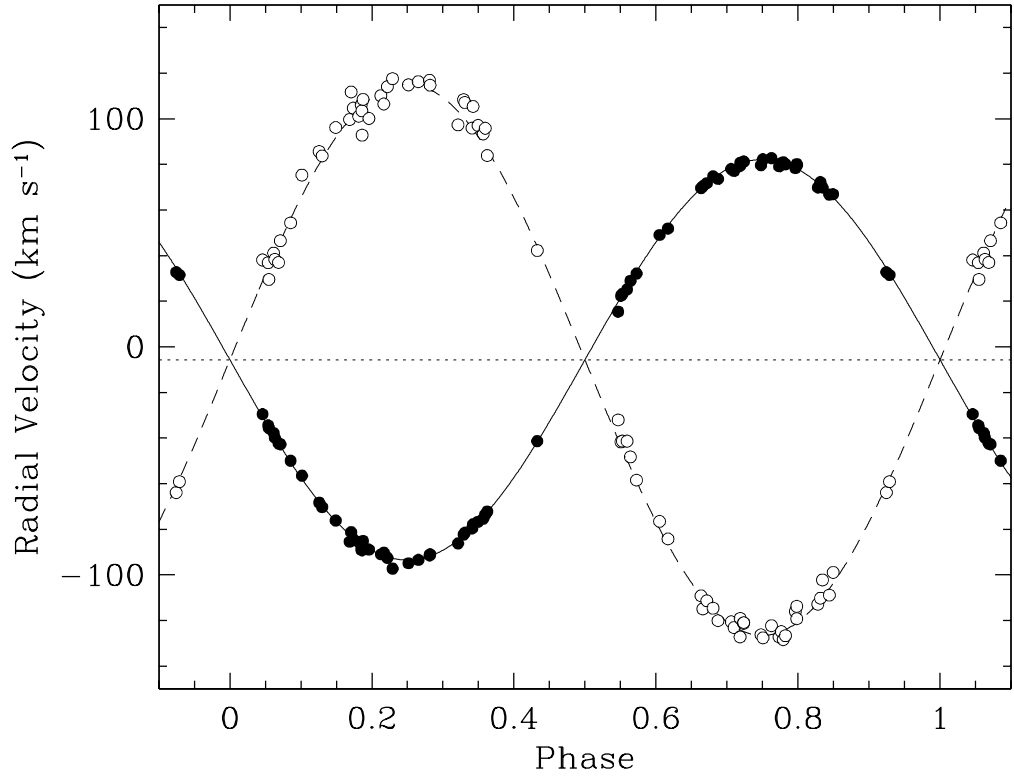}
\vskip -0.5in
 \figcaption[]{Radial velocity observations for stars Aa
(filled circles) and Ab (open circles) in the eclipsing binary, with
our fitted orbit. The motion of the binary in the wide orbit has been
subtracted. Phase 0.0 corresponds to the primary eclipse, and the
dotted line represents the center-of-mass velocity of the
system.\label{fig:rvbin}}
 \end{figure}

\begin{figure} 
\epsscale{0.90}
\vskip -0.1in
\plotone{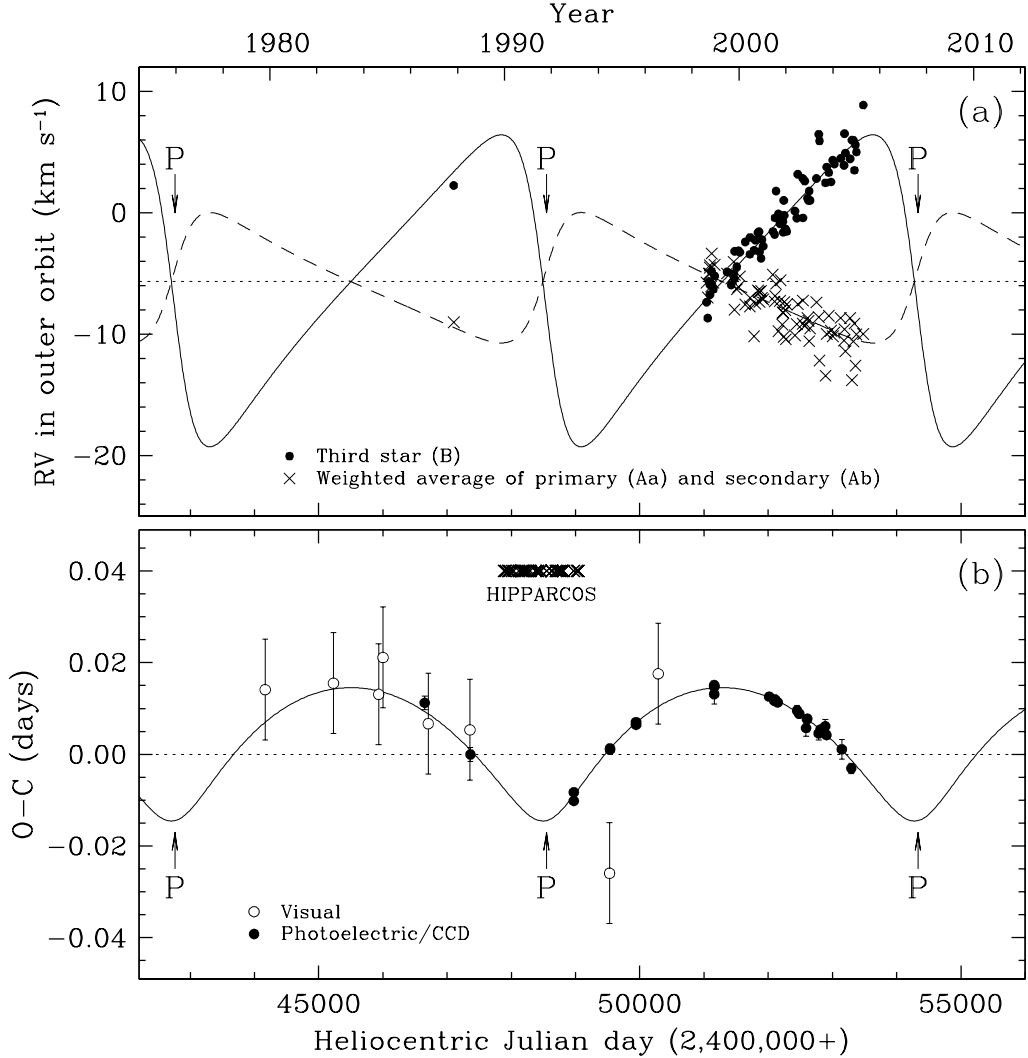}
\vskip 0.0in
 \figcaption[]{Radial velocities for \V10 in the outer orbit, and
$O-C$ timing residuals as a function of time. The curves represent our
best-fit model that uses velocities simultaneously with the eclipse
timings and HIPPARCOS observations. (a) Velocities for the third star
(filled circles) and the center-of-mass of the inner binary (crosses).
The latter is computed at each date from the weighted average of the
two stars, after removing the motion in the inner orbit.  The
center-of-mass velocity is indicated with the dotted line. (b) $O-C$
diagram of the more recent eclipse timings based on the linear
ephemeris in Table~\ref{tab:allelements}. Open circles are for visual
measurements, and filled circles for photoelectric/CCD
measurements. The older photographic timings have much larger error
bars and are not shown, for clarity. The dates of the HIPPARCOS
observations are indicated, and happen to cover periastron passage
(arrows).\label{fig:rvteromc}}
 \end{figure}

The radial velocity observations and orbital solution for the inner
eclipsing pair are shown graphically in Figure~\ref{fig:rvbin}, with
the motion in the outer orbit removed.  In Figure~\ref{fig:rvteromc}a
we show the velocities of the third star as a function of time, along
with the measured velocities of the center of mass of the inner binary
and the computed curves for the outer orbit.  The $O-C$ timing
residuals (primary and secondary) from the best-fit linear ephemeris
(Table~\ref{tab:allelements}) are displayed as a function of time in
Figure~\ref{fig:rvteromc}b, for the more recent visual and
photoelectric/CCD measurements.  The solid line represents the
computed third body effect. The semi-amplitude of this curve, as
derived from our global fit, is $K_{\rm O-C} = 0.01458 \pm 0.00048$
days. Residuals from the curve for the individual timings are listed
in Table~\ref{tab:timings}. The time axis in both panels of
Figure~\ref{fig:rvteromc} is the same to show the complementarity of
the data. Additionally, the time location of the HIPPARCOS
observations is shown in the lower panel, and happens to be centered
around periastron passage in the outer orbit. 

\begin{figure} 
\vskip 0in
\epsscale{0.87} 
\plotone{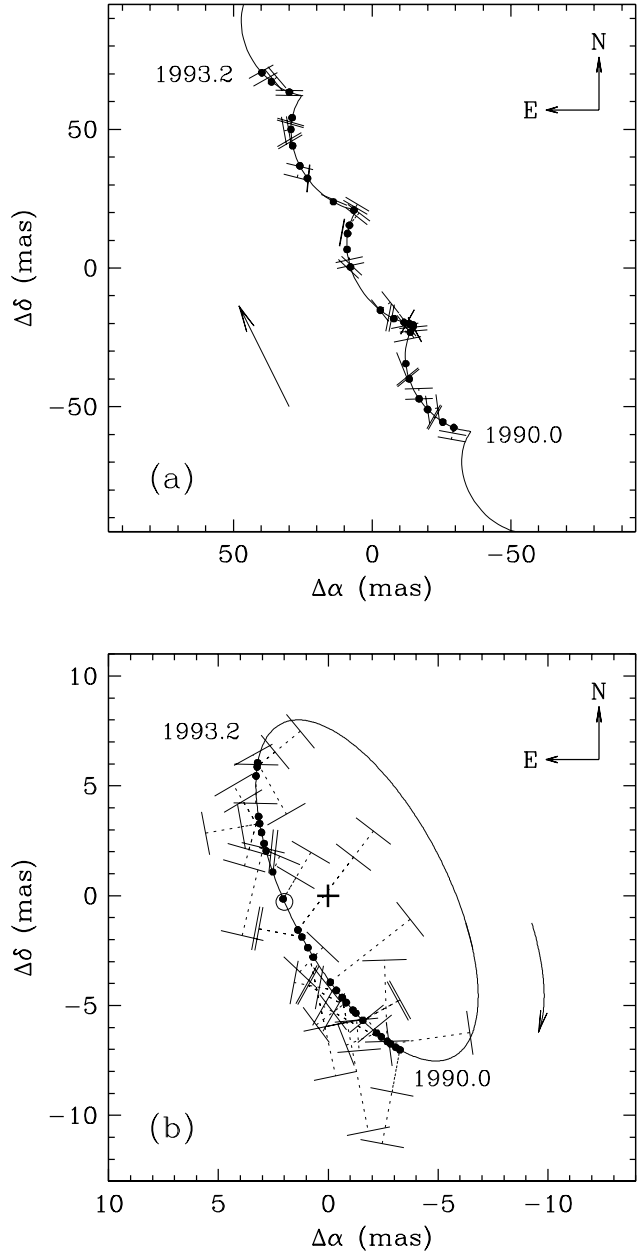}
\vskip 0.7in
 \figcaption[]{Path of the center of light of \V10 on
the plane of the sky, along with the HIPPARCOS observations (abscissae
residuals). See text for an explanation of the graphical
representation of these one-dimensional measurements.  (a) Total
motion resulting from the combined effects of parallax, proper motion,
and orbital motion according to the global solution described in the
text.  The arrow indicates the direction and magnitude of the annual
proper motion. (b) Residual orbital motion after subtracting the
parallactic and proper-motion contributions.  The plus sign represents
the center of mass of the triple, and the direction of motion in the
orbit (retrograde) is indicated by the arrow. Periastron is shown with
an open circle.\label{fig:hiporb}}
 \end{figure}

The astrometric motion of the photocenter of \V10 on the plane of the
sky is illustrated in Figure~\ref{fig:hiporb}, where the axes are
parallel to the Right Ascension and Declination directions. The top
panel shows the total motion resulting from the combined effects of
annual parallax, proper motion, and orbital motion. The dominant
contribution is from the proper motion (nearly 40~mas~yr$^{-1}$),
which is indicated with an arrow. Parallax and orbital motion are
smaller and comparable effects. In Figure~\ref{fig:hiporb}b we have
subtracted the proper motion and parallactic contributions, leaving
only the orbital motion with a 15.8-yr period and a semimajor axis of
9.6~mas. The direction of motion (retrograde) is indicated by the
arrow. The individual HIPPARCOS observations are depicted
schematically in both panels of this figure, but are seen more clearly
in Figure~\ref{fig:hiporb}b.  Because they are one-dimensional in
nature, the exact location of each measurement on the plane of the sky
cannot be shown graphically. The filled circles represent the
predicted location on the computed orbit. The dotted lines connecting
to each filled circle indicate the scanning direction of the HIPPARCOS
satellite for each measurement, and show which side of the orbit the
residual is on. The short line segments at the end of and
perpendicular to the dotted lines indicate the direction along which
the actual observation lies, although the precise location is
undetermined.  Occasionally more than one measurement was taken along
the same scanning direction, in which case two or more short line
segments appear on the same dotted lines.  Figure~\ref{fig:hiporb}b
shows that the observations are all on the long side of the orbit, and
due to its orientation, they are roughly parallel to the direction of
the proper motion. Since the orbit was not known at the time the
HIPPARCOS catalog was released, we expect the orbital motion to have
been absorbed into the published proper motion to some
extent. Although the effect is not large, we do see a hint of this in
that both components of the proper motion as published are larger than
the values we derive from our global solution, by about
2.5~mas~yr$^{-1}$ and 4.0~mas~yr$^{-1}$ for $\mu_{\alpha}^*$ and
$\mu_{\delta}$, respectively. Statistically the effect is somewhat
more significant in Declination.
	
\section{Absolute dimensions}
\label{sec:dimensions}

The combination of the photometric results in \S\ref{sec:photometry}
and the spectroscopic results in \S\ref{sec:orbit} yields the absolute
masses and radii of the components of the eclipsing binary, which we
list in Table~\ref{tab:physics} along with other properties as well as
inferred parameters for the third star (see \S\ref{sec:tertiary}). For
the photometric results we have adopted the weighted average of the
Gettysburg and URSA elements, which are $r_{\rm Aa} = 0.1668 \pm
0.0017$, $k = 0.598 \pm 0.004$, and $i_{\rm A} = 88\fdg2 \pm 0\fdg4$.
The precision of the absolute masses of the binary components is 1.2\%
for star Aa and 0.7\% for Ab, while that of star B is only 4\%. The
absolute radii are determined to 1\% and 2\% for the primary and
secondary, respectively.  The measured projected rotational velocities
of both stars are in excellent agreement with the values computed for
synchronous rotation, as expected for such a short period.

The parallax determination is much improved compared to the original
HIPPARCOS result, the error being reduced from 17\% to about 3\%
driven mostly by the individual estimates for stars Aa and Ab based on
their physical properties, and to some extent also by the astrometry.
The increased accuracy has an impact on the comparison with theory
described below. The corresponding distance to the system is $166.9
\pm 5.6$ pc.  The space motion of \V10 in the Galactic frame is $U =
-20$ \kms, $V = -2$ \kms, and $W = +16$ \kms\ (relative to the Local
Standard of Rest), quite typical of Population~I stars.  The
inclination angles of the inner and outer orbits ($i_{\rm A}$ and
$i_{\rm AB}$) are known from our spectroscopic and light curve
analyses, but the relative inclination $\phi$ between the two orbits
cannot be determined because the position angle of the node for the
eclipsing pair is unknown\footnote{The relative inclination angle of
the two orbits is given by $\cos\phi = \cos i_{\rm A} \cos i_{\rm AB}
+ \sin i_{\rm A} \sin i_{\rm AB} \cos(\Omega_{\rm A}-\Omega_{\rm AB})$
\citep[e.g.,][]{Fekel:81}. Since $\Omega_{\rm A}$ is unknown, we can
only set limits to $\cos(\Omega_{\rm A}-\Omega_{\rm AB})$ between $-1$
and +1, which leads to $i_{\rm A} - i_{\rm AB} \leq \phi \leq i_{\rm
A} + i_{\rm AB}$.}.  However, a lower limit of $\phi_{\rm min} =
20\fdg2 \pm 4\fdg9$ can be placed, which appears to exclude
coplanarity. The semimajor axis of the photocenter of the binary in
the wide orbit is only 9.55~mas, but the relative semimajor axis is
much larger (55.3~mas).  Considering that the orbit is eccentric and
therefore that the angular separation can be as large as 80~mas, the
third star is potentially resolvable with current techniques (e.g.,
speckle interferometry on 4-m class telescopes, or adaptive optics on
8--10-m class telescopes) despite being 2 magnitudes fainter than the
binary in the visible. Maximum separation should occur around the year
2014.

The central surface brightness parameter from the light curve
solutions constrains the difference between the effective temperatures
of the stars in the binary. To estimate the mean temperature of the
system we made use of absolute $U\!BV$ photometry outside of eclipse
available from the work of \cite{Lacy:92}, the $B_T$ and $V_T$
magnitudes from the Tycho-2 catalog \citep{Hog:00}, and the
$J\!H\!K_s$ magnitudes from the 2MASS catalog. Color indices from
various combinations of these magnitudes along with calibrations by
\cite{Alonso:96} and \cite{Ramirez:05} lead to a mean temperature
close to 6000~K, similar to our estimate in \S\ref{sec:photometry}.
Folding in the spectroscopic estimates we obtain average values for
the primary and secondary of $6180 \pm 100$~K and $5300 \pm 150$~K,
respectively, which we adopt here. These correspond to spectral types
of approximately F9 and G8, while the third star is roughly G4
\citep{Cox:00}.
	
\section{Discussion}
\label{sec:discussion}

The high precision of the mass and radius determinations for stars Aa
and Ab (errors of 1--2\%), along with the subsolar secondary mass and
a mass ratio that is quite different from unity ($q = 0.7266$), make
\V10 a particularly valuable system to test models of single-star
evolution. The four measured properties available for testing are the
mass, radius, effective temperature, and luminosity ($M_V$). The
latter two are of course related through the Stefan-Boltzmann law,
although they were determined independently here since we know the
distance. Because the temperatures were derived somewhat more
indirectly and have greater uncertainties than $M_V$, we have
preferred to rely here on the absolute visual magnitudes, inferred
from the apparent brightness of the system (ignoring extinction; see
\S\ref{sec:photometry}), the fractional luminosities, and the
parallax.

No accurate estimate of the metallicity is available. The Population~I
kinematics (\S\ref{sec:dimensions}) provide only circumstancial
evidence that the composition is perhaps near solar. In the following
we have made that assumption initially, although non-solar
compositions have also been explored by leaving the metallicity as a
free parameter. In principle this additional freedom may appear to
weaken the test; however, in this particular case it has no effect on
our main conclusion, as seen below.

\subsection{Comparison with evolutionary models}
\label{sec:models}

\begin{figure} 
\vskip 0in
\epsscale{1.0} 
\plotone{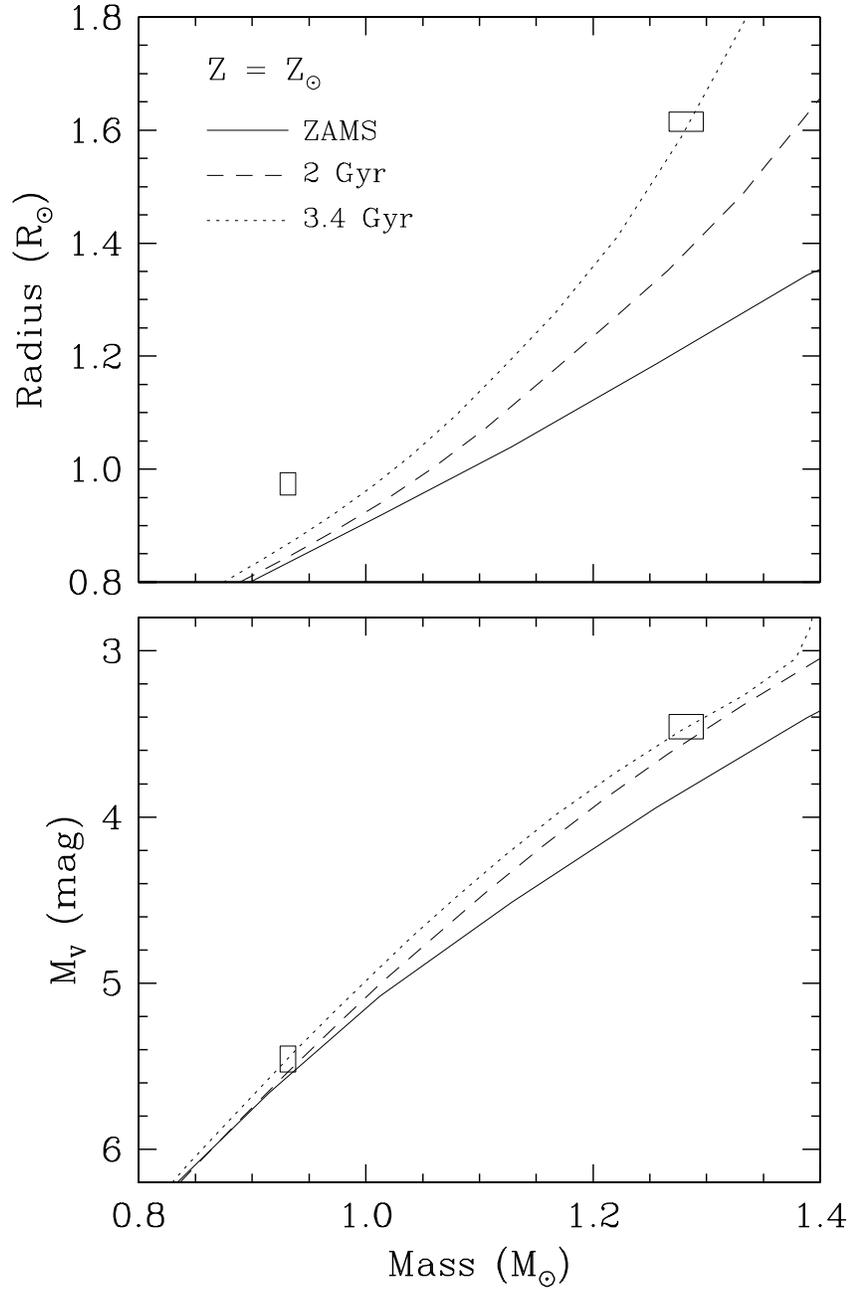}
\vskip 0.7in

 \figcaption[]{Comparison between the observed properties (error
boxes) of the primary and secondary of \V10 (mass, radius, absolute
visual magnitude) and model isochrones by \cite{Yi:01} and
\cite{Demarque:04} for solar metallicity and three different ages, as
labeled.\label{fig:yale}} \end{figure}

In Figure~\ref{fig:yale} we display the radius and absolute visual
magnitude of \V10 as a function of mass, along with theoretical
isochrones from the Yonsei-Yale series of evolutionary models
\citep{Yi:01, Demarque:04} for solar metallicity ($Z_{\sun} \equiv
0.01812$). As do most models, these treat convection in the
mixing-length approximation, and adopt a mixing-length parameter that
best fits the properties of the Sun ($\alpha_{\rm ML} = 1.7432$).  The
color transformations and bolometric corrections adopted are from the
tables by \cite{Lejeune:98}.  The figure shows that, within the
errors, a 3.4-Gyr isochrone for this metallicity provides an excellent
match to all observations except for the radius of the secondary,
which is $\sim$10\% larger than predicted. This very significant
difference is 5 times the uncertainty. Since the luminosity of the
secondary is well reproduced by the models, it follows that the
effective temperature must be cooler than predicted. Indeed, our
estimate from \S\ref{sec:dimensions} is $T_{\rm eff}^{\rm Ab} = 5300
\pm 150$~K, formally cooler than the predicted value of 5460~K but
only at the 1$\sigma$ level. A more precise estimate of the secondary
temperature may be derived from our $M_V$ and the radius of the star
\citep[with bolometric corrections taken also from][for
consistency]{Lejeune:98}.  This gives $T_{\rm eff}^{\rm Ab} = 5210 \pm
80$~K, which is lower than the models predict at a more significant
3.1$\sigma$ level. Thus the secondary does appear to be too cool by
roughly 200 K.

\begin{figure} 
\vskip -0.2in
\epsscale{0.80} 
\plotone{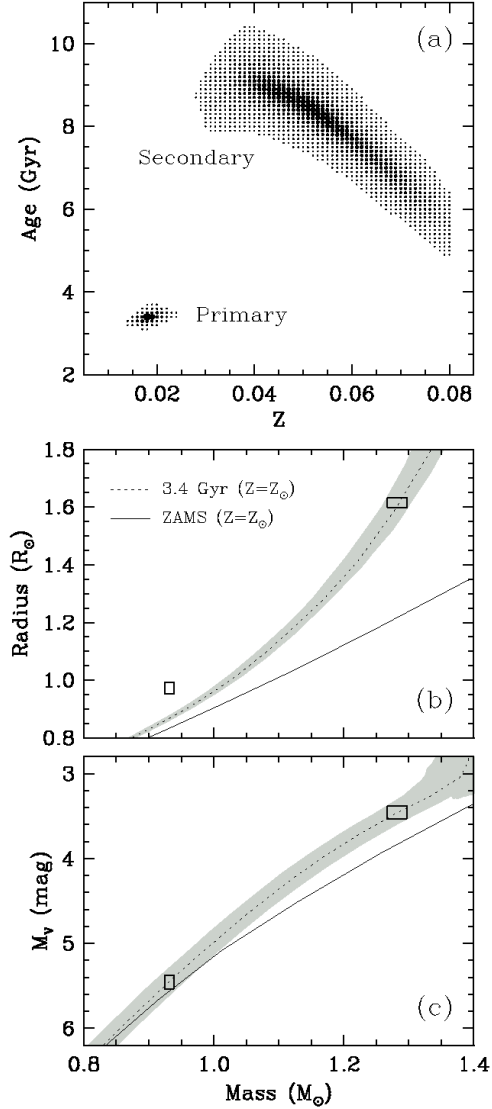}
\vskip -0.4in

 \figcaption[]{Age and metallicity range of theoretical models by
 \cite{Yi:01} and \cite{Demarque:04} that fit the measured properties
 of the primary and secondary components of \V10. (a) Shaded areas
 represent age-metallicity combinations consistent with the measured
 mass, radius, and absolute visual magnitude of each star.  The
 greyscale level is a measure of how well each model matches the exact
 values of $M$, $R$ and $M_V$ (darker areas represent a better fit);
 (b) The shaded area represents the projection onto the mass/radius
 diagram of all models indicated in the top panel that fit the primary
 star within its error boxes. None of the models come close to
 reproducing the secondary radius; (c) Same as above, for the
 mass/$M_V$ plane. The secondary star is well fit by the same models
 that match the primary. \label{fig:metage}}

 \end{figure}

Extensive tests show that adjustments in the age or metallicity of the
isochrones do not improve the agreement. Figure~\ref{fig:metage}a
illustrates this in the age/metallicity plane. The shaded areas
represent all isochrones that are consistent with the measured mass,
radius, and $M_V$ of each star within the estimated uncertainties.
The shaded areas do not overlap, meaning that no single isochrone can
reproduce all properties of both stars simultaneously. The region
spanned by models that are consistent with the primary properties
([Fe/H] within $\pm$0.14 dex of solar, ages of 3.06--3.78 Gyr) are
shown in the mass/radius and mass/$M_V$ diagrams in
Figures~\ref{fig:metage}b and \ref{fig:metage}c. None of those models
come close to matching the radius of the secondary star, although they
do agree well with its absolute magnitude, as stated above.

As a test we repeated the comparison using a different set of
isochrones by \cite{Girardi:00}, from the Padova group. In these
models the adopted mixing-length parameter that best reproduces the
Sun (for $Z_{\sun} \equiv 0.019$) is $\alpha_{\rm ML} = 1.68$, and a
number of other physical ingredients are somewhat different, as are
the color transformations and bolometric corrections. Nevertheless,
the fits to \V10 are very similar, and again indicate a large
discrepancy in the secondary radius $R_{\rm Ab}$ and a smaller
difference in temperature.

Both of the above sets of evolutionary models adopt a grey
approximation for the outer boundary conditions to the internal
stucture equations. More sophisticated non-grey atmospheres have been
used in the models by \cite{Siess:97} and \cite{Baraffe:98}, which has
been shown to be important for low mass stars \citep[see,
e.g.,][]{Chabrier:97}.  Additionally, these models incorporate an
improved equation of state and other refinements, although all this
should only have an impact for stars considerably smaller than the
secondary of \V10 (which is near the calibration point at
1~M$_{\sun}$). Indeed, the comparison between \V10 and the
\cite{Baraffe:98} isochrones ($Z_{\sun} \equiv 0.020$, $\alpha_{\rm
ML} = 1.9$) is not very different from the two previous fits, and
yields a similar age as before based on the agreement with all
observed properties except $R_{\rm Ab}$.

The measured radii depend mainly on the results from the light curve
fits and to a lesser degree on the spectroscopy.  As described
earlier, tests carried out during the analysis of both types of data
make it very unlikely that either set of results is biased enough to
explain the discrepancy.  Since the properties of the more massive
primary star appear to be well reproduced by the models \citep[as
expected from previous experience; see][]{Andersen:03}, suspicion
falls on our theoretical understanding of stars under a solar mass.
Similar evidence has been presented over the last several years
\citep[e.g.,][]{Popper:97, Clausen:99a, Torres:02, Ribas:03,
Dawson:04, Lopez-Morales:05}, but in fact indications go as far back
as the work by \cite{Hoxie:73}, \cite{Lacy:77}, and others. All of
these studies have shown that theoretical calculations for stars less
massive than the Sun tend to underestimate the radius by as much as
10\% to 20\%. Temperature differences also appear to be present, as in
\V10.

\begin{figure} 
\vskip -0.3in
\epsscale{1.0} 
\plotone{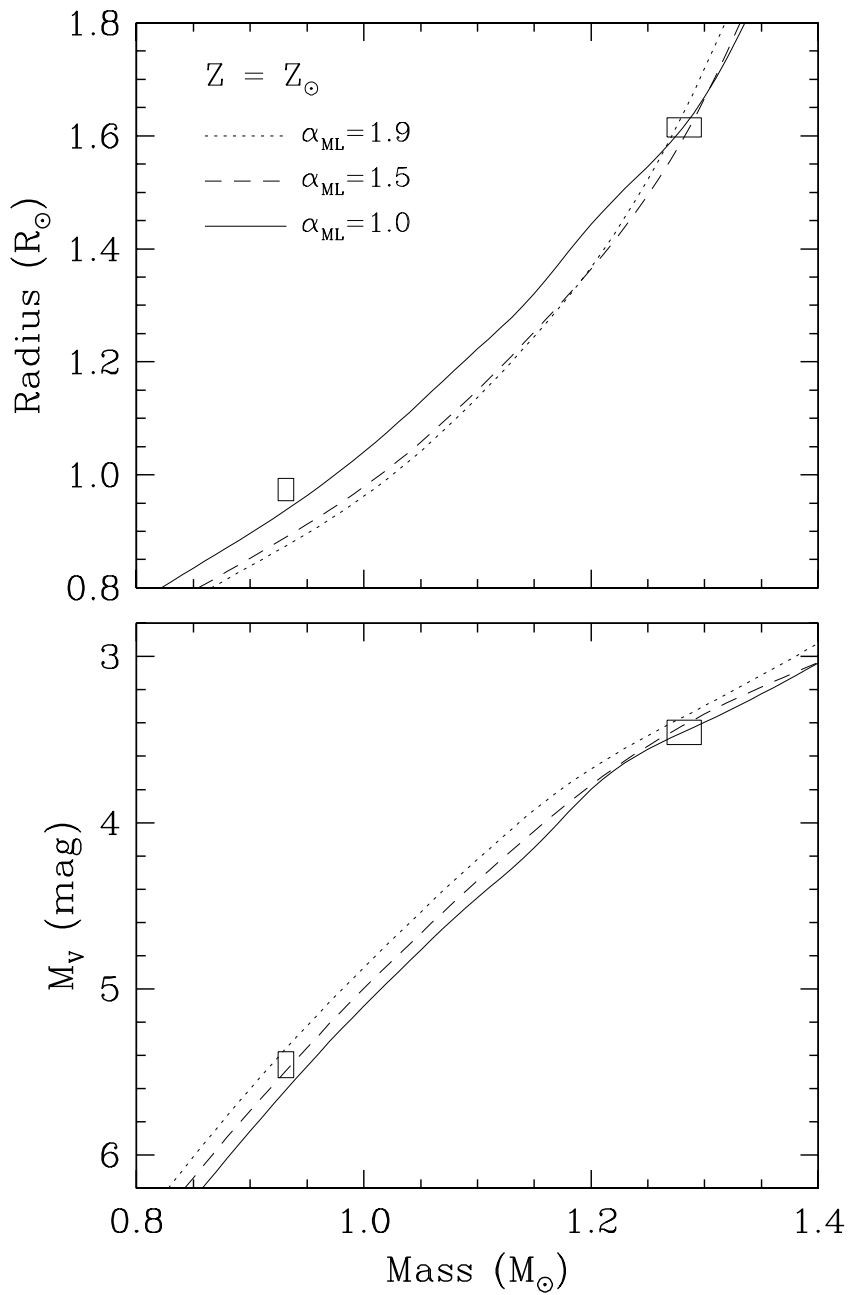}
\vskip 0.4in

 \figcaption[]{Observations of \V10 (error boxes) compared against
 model isochrones by \cite{Baraffe:98} for three different values of
 the mixing length parameter, as labeled.  Solar metallicity is
 assumed, and the age of each isochrone has been tuned to fit the
 properties of the primary as closely as possible within errors (2.6
 Gyr for $\alpha_{\rm ML} = 1.0$, 3.1 Gyr for $\alpha_{\rm ML} = 1.5$,
 and 3.4 Gyr for $\alpha_{\rm ML} = 1.9$). The model with $\alpha_{\rm
 ML} = 1.0$ (solid line) is seen to provide a much closer match to the
 radius of the secondary than the one with $\alpha_{\rm ML}$ adjusted
 to fit the Sun. \label{fig:baraffe}}
 \end{figure}

The models by \cite{Baraffe:98} we used above are also available for
lower values of the mixing length parameter in the mass range of
interest. In Figure~\ref{fig:baraffe} we compare the observations for
\V10 against solar metallicity models for $\alpha_{\rm ML} = 1.0$ and
1.5, in addition to $\alpha_{\rm ML} = 1.9$, adjusting the age to
satisfy the constraints on the primary star within the errors.  A
2.6-Gyr isochrone with $\alpha_{\rm ML} = 1.0$ provides a
significantly better fit to the secondary radius than the one with
$\alpha_{\rm ML} = 1.9$ used earlier, although the model prediction
still falls short of the observed $R_{\rm Ab}$ value by about
1.8$\sigma$ at the measured mass.  The reduced convection implied by
the lower mixing length parameter in these models compared to the
previous ones has the effect of increasing the theoretical radii and
slightly lowering the effective temperatures. The predicted
temperature for $\alpha_{\rm ML} = 1.0$ and $M = M_{\rm Ab}$ is 5110
K, somewhat closer than before to the empirical result of 5210 K (see
above).

The considerably smaller value of $\alpha_{\rm ML}$ apparently
required by a star so close to the solar mass as \V10 Ab is perhaps
somewhat surprising, although other studies have found similar
evidence in some stars and have even argued for a mass dependence,
with smaller values of $\alpha_{\rm ML}$ for later-type stars
\citep[e.g.,][]{Lastennet:03}.  However, the observational evidence
for this is often contradictory \citep[see][]{Eggenberger:04}.
Theoretical studies, on the other hand, have tended to predict the
opposite dependence with mass \citep{Ludwig:99, Trampedach:99}.

\subsection{Comparison with other stars: the activity-radius connection}
\label{sec:activity}

Further clues on these disagreements may be gained from looking at
other eclipsing binary systems with well determined properties. To
avoid any possible dependence on mass, we focus here on binaries
having at least one star in the same mass range as the secondary of
\V10. Three such systems are available with reliable determinations of
the mass, radius, and temperature good to better than 3\%: RW~Lac
\citep{Lacy:05}, in which the primary is within 0.5\% of the mass of
\V10~Ab; FL~Lyr, with a secondary only 3\% more massive; and HS~Aur,
with a primary that is 3.5\% less massive. The latter two are from
work by \cite{Popper:86} and \cite{Andersen:91}, and FL~Lyr has the
added advantage that the components have quite dissimilar masses (as
in the case of \V10), which provides increased leverage for the
test. The main properties of these systems are collected in
Table~\ref{tab:otherstars}, along with \V10 and the Sun for reference.
Two additional systems \citep[CG~Cyg and RT~And;][]{Popper:94} have
one component in the same mass range, but the temperature
determinations are considerably more uncertain and are thus less
useful for our purposes.

\begin{figure} 
\vskip -0.3in
\epsscale{0.9} 
\plotone{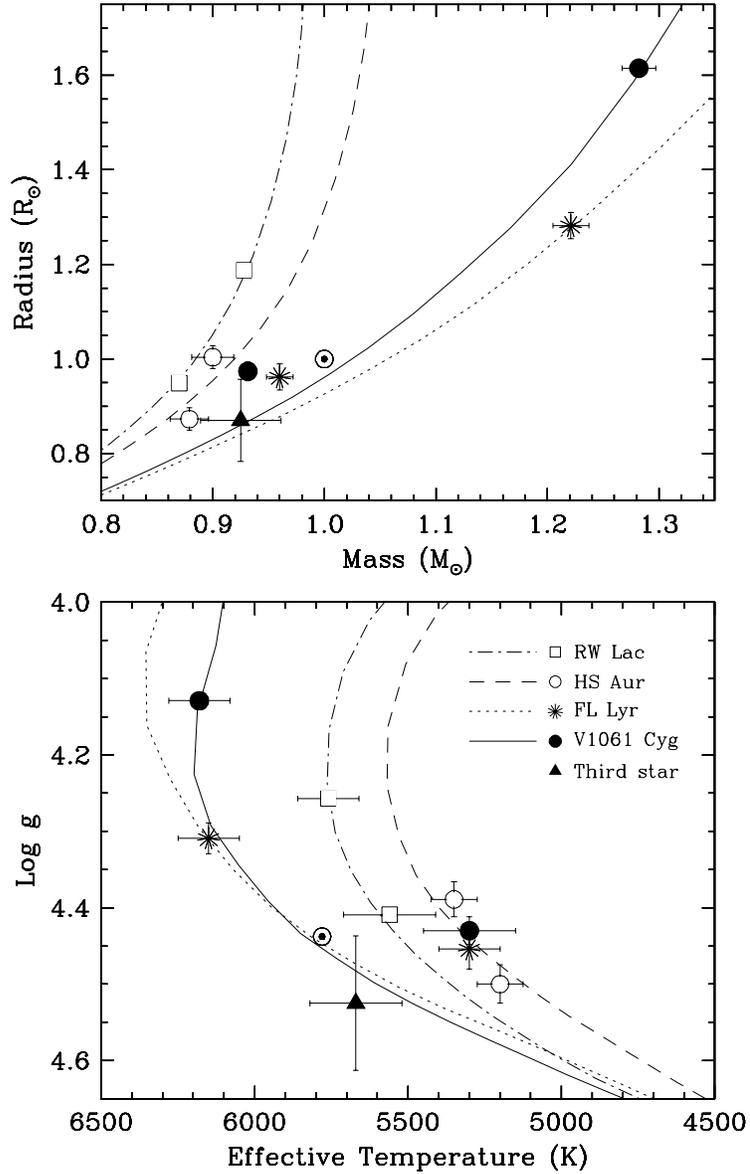}
\vskip 0.6in 

 \figcaption[]{Properties of \V10 and three other eclipsing binaries
having one star closely matching \V10 Ab in mass, compared against the
Yonsei-Yale models by \cite{Yi:01} and \cite{Demarque:04}.  For each
binary system the model isochrone best fitting all observed properties
is shown: $Z = 0.0118$ and age = 10.7~Gyr for RW~Lac, $Z = 0.024$ and
age = 10.5~Gyr for HS~Aur, $Z = 0.026$ and age = 2.4~Gyr for FL~Lyr,
and $Z = Z_{\odot}$ and age = 3.4~Gyr for \V10.  The secondary star in
FL~Lyr is significantly larger than predicted, as is \V10 Ab (see
text). The Sun ($\sun$) is also shown for reference, along with the
third component of \V10 (\S\ref{sec:tertiary}). \label{fig:otherstars}}

\end{figure}

We compared the measurements for RW~Lac, FL~Lyr, and HS~Aur to the
Yonsei-Yale models in the same way we did for V1061 Cyg, adjusting the
age and metallicity of the isochrones to obtain the best possible fit
(see Figure~\ref{fig:otherstars}). FL~Lyr is seen to present the same
problem as \V10: the secondary appears too large compared to theory,
by about the same amount as we saw before ($\sim$10\%), while the
properties of primary are well fit. On the other hand, the primary of
RW~Lac is virtually identical in mass to \V10 Ab yet it shows no
indication of a radius discrepancy. The primary in HS~Aur also seems
to be well reproduced by the models (as well as the secondary), within
the errors.

A pattern that may explain why stars of very similar mass sometimes
appear too large, while other times they conform well to theory, is
seen in the activity level they present. Both \V10 Ab and FL~Lyr B,
which show the radius discrepancy, are in relatively tight systems
with orbital periods of only 2.35 days and 2.18 days, respectively.
The stars in these binaries are rapid rotators (consistent with
synchronous rotation maintained by tidal forces): \V10 Ab has a
measured $v \sin i = 20 \pm 3$~\kms\ (\S\ref{sec:spectroscopy}) and
FL~Lyr B has $v \sin i = 25 \pm 2$~\kms\ \citep{Popper:86}. Both
binaries are strong ROSAT X-ray sources \citep{Voges:99} and their
X-ray luminosities (Table~\ref{tab:otherstars}) are in line with those
of active single and binary stars with similar $v \sin i$ values
\citep[e.g.,][Fig.\ 5]{Cutispoto:03}. Although to our knowledge no
observations exist to verify whether the \ion{Ca}{2} H and K lines are
in emission, both systems show signs of intrinsic variability in the
light curve suggesting the presence of spots (see Popper et al.\ 1986
for FL~Lyr, and \S\ref{sec:photometry} for \V10). They are thus active
binaries.  RW~Lac and HS~Aur, on the other hand, do not show a radius
discrepancy and happen to have much longer orbital periods of 10.37
days and 9.82 days, respectively, and their components are slow
rotators. No \ion{Ca}{2} emission is seen in HS~Aur
\citep{Popper:76}. Neither binary was detected as an X-ray source by
ROSAT \citep{Voges:99}, and by all accounts they appear inactive.  A
direct relation is thus seen for stars of the same mass between the
activity level and the increased stellar size compared to predictions
from standard models (i.e., those adopting a mixing length parameter
matching the Sun): \emph{active stars are larger, and inactive ones
appear normal}. The evidence for \V10 and other later-type stars also
indicates that active stars are cooler.

This seems to have been mentioned only as a possibility (among several
others) in previous studies reporting radius discrepancies for stars
under a solar mass, at least those originating from the eclipsing
binary community.  But as a matter of fact, the connection between
activity and the global properties of low mass stars has been studied
in some detail previously in a slightly different context, although
without direct and precise knowledge of the masses and radii of the
stars involved.  \cite{Mullan:01} investigated the effects of magnetic
fields on the sizes and effective temperatures of active versus
inactive M dwarfs, and found empirical evidence that a higher activity
level leads to larger radii and cooler temperatures.  Their sample
consisted of single stars with effective temperature determinations
from infrared spectroscopy and bolometric luminosities from multi-band
photometry, from which stellar radii were inferred indirectly with
typical uncertainties of 10--15\%.  Although their work focussed
mainly on the consequences for the internal structure of fully
convective stars, their initial attempts at modeling magnetic fields
were successful in describing these effects to first order.  The
present study on \V10 shows the activity-radius connection clearly for
stars with accurately determined dynamical masses, radii, and
temperatures good to 1--3\%. More importantly, the effect is seen for
objects that are only 7\% less massive than the Sun.

A theoretical understanding of this connection is also not new, and
provides some insight into the better agreement between the
observations for \V10 and the low-$\alpha_{\rm ML}$ models in
Figure~\ref{fig:baraffe}.  Strong magnetic fields commonly associated
with chromospheric activity have been shown to inhibit the efficiency
of convective heat transport \citep[e.g.,][and references
therein]{Bray:64, Gough:66, Stein:92}, and as a result the size of the
star must grow larger to radiate away the same amount of energy. The
decreased convection effectively leads to a lower value of the mixing
length parameter \citep[see][]{Tayler:87}, which explains the better
fit to the radius of the secondary using the \cite{Baraffe:98} models
with $\alpha_{\rm ML} = 1.0$ as opposed to those with the solar value
of $\alpha_{\rm ML} = 1.9$.  Similar improvements in the fit to other
low-mass stars using a reduced mixing length parameter were also
reported by \cite{Clausen:99a}, and in fact much earlier by
\cite{Gabriel:69} and \cite{Cox:81}.

\begin{figure} 
\vskip 0.2in 
\epsscale{0.90} 
\plotone{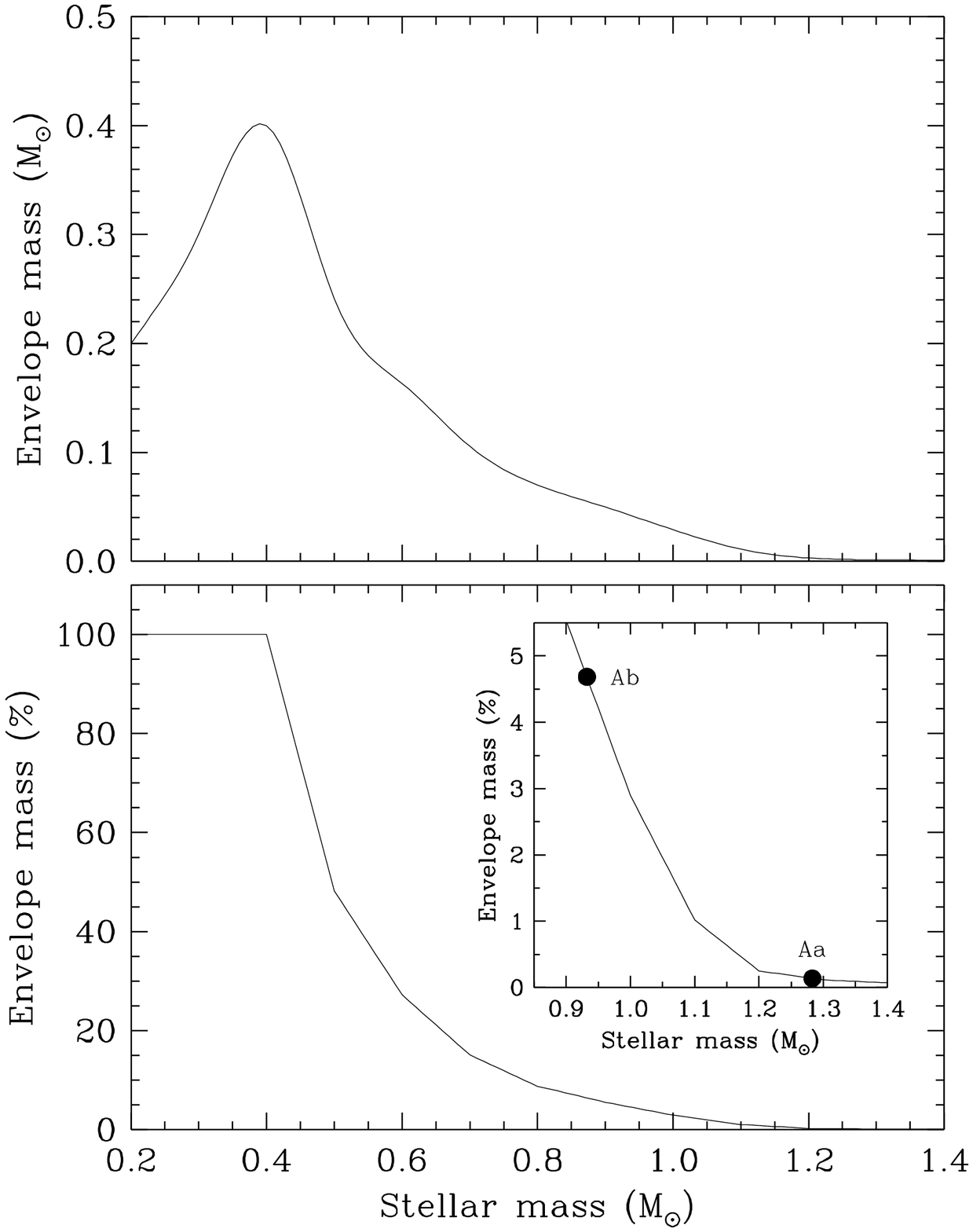}
\vskip 0.5in
 \figcaption[]{Mass of the convective envelope for
main-sequence stars as a function of stellar mass, based on evolution
models by \cite{Siess:97} and \cite{Siess:00} for solar composition
and the estimated age for \V10 of 3.4 Gyr. The envelope mass is shown
in solar masses (top) and as a fraction of the stellar mass (bottom).
According to these models stars of 0.4~M$_{\sun}$ and lower are fully
convective. In the inset the region around the masses of \V10 is shown
on a larger scale, and the envelope mass of both stars is indicated.
\label{fig:envelope}}
 \end{figure}

The enhanced activity in \V10 Ab is driven by rapid (synchronous)
rotation induced by tidal forces. The primary star is rotating even
more rapidly ($v \sin i = 36 \pm 2$~\kms), yet it shows no obvious
indication of a significantly larger radius compared to standard
models. This is most likely because it is a more massive star
(spectral type F9, $M = 1.282$~M$_{\sun}$) and therefore its
convective envelope (where magnetic activity takes place) is
significantly reduced. To illustrate this in a more quantitative way,
Figure~\ref{fig:envelope} shows the mass of the convective envelope
for main-sequence stars as a function of stellar mass from the models
by \cite{Siess:97} \citep[see also][] {Siess:00}, for solar
composition and the age of 3.4~Gyr we infer for the system. In the
lower panel the envelope mass is expressed as a percentage of the
stellar mass. The inset shows an enlargement of the region relevant to
\V10. According to these models the convective envelope of the
secondary represents about 4.7\% of its total mass, whereas that of
the primary is only 0.2\% ($\sim$20 times smaller). A similarly
reduced convective envelope for the primary of FL~Lyr ($M =
1.221$~M$_{\sun}$) explains why that star also does not show evidence
of an enlarged radius, while the secondary (which is similar in mass
to \V10 Ab) does.

The extent to which a star is enlarged by this effect may be expected
to depend upon the strength of the activity. We note, for instance,
that although models with $\alpha_{\rm ML} = 1.0$ seem to match the
radius of \V10 Ab better, the same models still do not reproduce the
properties of apparently more active lower-mass stars such as YY~Gem
\citep{Torres:02}, CU~Cnc \citep{Ribas:03}, GU~Boo
\citep{Lopez-Morales:05}, and others.  It is clear that further
examples of stars (both active and inactive) in the suitable mass
range with accurately determined parameters are needed to explore
this.  Systems such as \V10 and FL~Lyr, with components of appreciably
different mass and primaries that are larger than
$\sim$1.2~M$_{\sun}$, are particularly useful because they allow the
activity-radius effect to be separated out, since the primary should
not be significantly affected.

\subsection{The tertiary star}
\label{sec:tertiary}

The \V10 system is especially valuable for containing a third star of
mass indistinguishable from that of the active star Ab, but with slow
rotation. It is presumably an inactive analog of the secondary, with
the same age and chemical composition, and is thus ideal for testing
the effect discussed above. We do not have a direct measurement of its
size, but the radius can be inferred from the effective temperature
and luminosity, and is $R_{\rm B} = 0.870 \pm 0.087$~R$_{\sun}$.
While this is formally smaller than the secondary radius of $R_{\rm
Ab} = 0.974 \pm 0.020$~R$_{\sun}$, as we expected, it is not precise
enough for a definitive test. Other inferred properties of the
tertiary are listed in Table~\ref{tab:physics}. The star is
represented graphically in Figure~\ref{fig:otherstars}, where it is
seen to be in good agreement with the isochrone for the primary,
within the admittedly large errors.  The effective temperature we
determine for the tertiary ($5670 \pm 150$ K) is somewhat hotter than
that of the secondary even though the masses are similar, which again
is consistent with our conclusion that the activity in the secondary
has made that star cooler.

The formal uncertainty in $R_{\rm B}$ is dominated by errors in the
light fraction $\ell_{\rm B}$ and effective temperature $T_{\rm
eff}^{\rm B}$, in that order of importance. Improvements in the former
could be made, for example, by direct detection of the third star and
measurement of its brightness through speckle interferometry or
adaptive optics imaging (\S\ref{sec:dimensions}). A better temperature
estimate would require spectroscopy with higher S/N, although the
triple-lined nature of the spectrum and the faintness of the star
would still pose a challenge. Additional radial velocity measurements
over the coming years covering the next nodal passage in the outer
orbit (in 2009.1) would also be important to refine the value of the
tertiary mass.
		
\section{Concluding remarks}

The results of our spectroscopic, photometric, and astrometric
analyses of the \V10 system have taken us in a rather different
direction than we anticipated when we began this study. The possible
status of the object as an example of the rare class of cool Algols is
now clearly ruled out.  Instead, we have shown here that it is a
hierarchical triple system with an outer period of 15.8~yr, in which
the eclipsing inner pair is well detached, is composed of
main-sequence stars, has a mass ratio quite different from unity, and
has a secondary that is slightly below a solar mass. The absolute
masses and radii for the binary components are determined with a
relative precision of 2\% or better, and the mass of the third star is
good to 4\%.

While the primary star is well fit in mass, radius, temperature, and
luminosity by standard stellar evolution models with a metallicity
near solar and a mixing length parameter set by the calibration to the
Sun, the secondary appears $\sim$10\% too large. This discrepancy is 5
times the size of the observational errors, and quite surprising for a
star that differs by only 7\% in mass from the Sun. There are also
indications that it is cooler than predicted by some 200 K. \V10 is
yet another example highlighting our incomplete understanding of the
structure and evolution of stars in the lower main sequence.  Similar
differences in size and temperature have been noticed previously for
lower-mass stars with accurately measured properties, but the source
of the problem has remained unclear. By comparing \V10 Ab to several
other objects of nearly the same mass we have identified the activity
level as a key factor distinguishing cases that show the radius
discrepancy from those that do not. This link between activity and
increased radius has been mentioned in the literature before, but is
shown here for the first time for stars with accurately known masses,
radii, and temperatures.

It is often stated that the structure and evolution of stars
(particularly those close to a solar mass) are completely determined
once the chemical composition and mass are specified. It is quite
clear now that for stars of order 1~M$_{\sun}$ or less an additional
parameter must be taken into account, which has to do with the level
of chromospheric activity.  Whether this parameter is directly the
rotational speed (e.g., $v \sin i$) or period, the magnetic field
strength, the Rossby number \citep[e.g.,][]{Noyes:84, Basri:87}, or
some other more complicated activity indicator remains to be
determined.  To first order it appears that the effective mixing
length may be a useful proxy, but with one exception current stellar
evolution models are not publicly available for more than one value of
$\alpha_{\rm ML}$, so testing this is somewhat difficult in practice.
If the predictions from the models are to reach an accuracy matching
current observations of low-mass stars ($\sim$1--2\% relative errors
in the masses and radii), this effect can no longer be ignored.
Further progress will require more examples of binary systems with
well determined physical parameters and different levels of activity
in the relevant mass regime in order to help calibrate any such
parameter.

Testing models of single-star evolution by means of eclipsing
binaries, as astronomers have done for decades, might perhaps be seen
as part of the problem since the enhanced activity displayed by many
of these systems is a direct result of tidal synchronization that
occurs only in close binaries.  Although wider eclipsing pairs with
inactive components under 1~M$_{\sun}$ certainly do exist, they are
less common and in some respects more difficult to study. Stars near
the bottom of the main sequence, M dwarfs in particular, are found to
be active more often than not. 

The effects of magnetic fields on the evolution of stars have already
begun to be explored by theorists \citep[e.g.,][although the latter
authors focus on more massive stars]{D'Antona:00, Mullan:01,
Maeder:03}, and initial comparisons with observations are encouraging.
One area where this is likely to have a significant impact is the
study of T~Tauri stars, which are typically very active. Even in
substellar objects such as brown dwarfs activity appears to be quite
common, and might be expected to have similar consequences on their
structure. Some evidence for this has already been reported
\citep{Mohanty:04}.

\acknowledgements 

We are grateful to J.\ Andersen, P.\ Berlind, M.\ Calkins, J.\ Caruso,
D.\ W.\ Latham, R.\ P.\ Stefanik, and J.\ Zajac for their efforts at
the telescope to obtain the majority of the spectroscopic observations for
\V10 used in this work, and to R.\ J.\ Davis for maintaining the CfA
echelle database. The referee, J.\ Andersen, is thanked for a number
of insightful comments and suggestions that improved the original
manuscript.  We also thank J.\ M.\ Kreiner for providing unpublished
times of eclipse for the binary, as well as R.\ Neuh\"auser and B.\
Stelzer for assistance with the X-ray observations.  GT acknowledges
partial support for this work from NSF grant AST-0406183 and NASA's
MASSIF SIM Key Project (BLF57-04). LM and HS were supported by
Gettysburg College and the Delaware Space Grant Consortium. Additional
thanks go to Peter Mack and Gary Hummer for technical support at the
Gettysburg College Observatory.  This research has made use of the
SIMBAD database, operated at CDS, Strasbourg, France, and of NASA's
Astrophysics Data System Abstract Service. This work makes use of data
products from the Two Micron All Sky Survey, which is a joint project
of the University of Massachusetts and the Infrared Processing and
Analysis Center/California Institute of Technology, funded by NASA and
the NSF.

\clearpage

\begin{deluxetable}{ccccccccccc}
\tabletypesize{\scriptsize}
\tablecolumns{11
\tablewidth{0pt}
\tablecaption{Radial velocity measurements for \V10.\label{tab:rvs}}
\tablehead{\colhead{HJD}          & \colhead{}     & \colhead{RV$_{\rm Aa}$} & \colhead{RV$_{\rm Ab}$} & \colhead{RV$_{\rm B}$} & \colhead{(O$-$C)$_{\rm Aa}$} & \colhead{(O$-$C)$_{\rm Ab}$} & \colhead{(O$-$C)$_{\rm B}$} & \colhead{Inner} & \colhead{Outer} & \colhead{$\Delta T$\tablenotemark{b}} \\
           \colhead{(2,400,000+)} & \colhead{Year} & \colhead{($\kms$)}      & \colhead{($\kms$)}      & \colhead{($\kms$)}     & \colhead{($\kms$)}           & \colhead{($\kms$)}           & \colhead{($\kms$)}          & \colhead{Phase\tablenotemark{a}} & \colhead{Phase\tablenotemark{a}} & \colhead{(days)}}}
\startdata 
  47101.6142 &  1987.8347 & \phn$+$66.13 & $-$106.02    &   $+$2.24 &  $-$0.13 &  \phn$+$7.64  &  $-$1.13   &    0.8346 &  0.7504 & $-$0.0045 \\
  51039.8277 &  1998.6169 & \phn$-$35.07 & \phn$+$30.15 &   $-$7.37 &  $-$0.30 &  \phn$-$5.64  &  $-$0.27   &    0.0549 &  0.4313 & $-$0.0143 \\
  51057.7385 &  1998.6660 & \phn$+$74.19 & $-$119.55    &   $-$8.69 &  $-$1.80 &  \phn$-$2.80  &  $-$1.69   &    0.6874 &  0.4343 & $-$0.0143 \\
  51075.7492 &  1998.7153 & \phn$-$71.66 & \phn$+$84.41 &   $-$5.67 &  $+$0.32 &  \phn$-$2.39  &  $+$1.22   &    0.3624 &  0.4375 & $-$0.0144 \\
  51087.6477 &  1998.7478 & \phn$-$40.88 & \phn$+$42.78 &   $-$5.91 &  $+$0.30 &  \phn$-$1.56  &  $+$0.91   &    0.4328 &  0.4395 & $-$0.0144 \\
  51089.6402 &  1998.7533 & \phn$-$90.48 & $+$115.25    &   $-$6.76 &  $+$0.79 &  \phn$+$1.99  &  $+$0.05   &    0.2819 &  0.4399 & $-$0.0144 \\
  51092.6176 &  1998.7614 & \phn$+$22.87 & \phn$-$41.20 &   $-$5.72 &  $+$0.59 &  \phn$+$1.83  &  $+$1.07   &    0.5507 &  0.4404 & $-$0.0144 \\
  51114.6251 &  1998.8217 & \phn$+$31.92 & \phn$-$58.71 &   $-$4.83 &  $-$0.77 &  \phn$-$1.23  &  $+$1.84   &    0.9289 &  0.4442 & $-$0.0144 \\
  51122.6356 &  1998.8436 & \phn$-$77.41 & $+$105.84    &   $-$5.94 &  $+$1.28 &     $+$10.08  &  $+$0.68   &    0.3425 &  0.4456 & $-$0.0144 \\
  51146.6110 &  1998.9093 & \phn$+$25.53 & \phn$-$40.99 &   $-$6.31 &  $-$1.10 &  \phn$+$8.35  &  $+$0.17   &    0.5593 &  0.4497 & $-$0.0145 \\
  51163.5148 &  1998.9556 & \phn$+$83.04 & $-$121.98    &   $-$5.20 &  $+$0.87 &  \phn$+$3.89  &  $+$1.18   &    0.7627 &  0.4526 & $-$0.0145 \\
  51356.6644 &  1999.4844 & \phn$-$42.89 & \phn$+$46.35 &   $-$4.87 &  $+$0.90 &  \phn$-$0.04  &  $+$0.40   &    0.0711 &  0.4860 & $-$0.0145 \\
  51423.5273 &  1999.6674 & \phn$+$28.65 & \phn$-$48.61 &   $-$5.95 &  $+$0.28 &  \phn$+$4.70  &  $-$1.06   &    0.5640 &  0.4976 & $-$0.0145 \\
  51464.6818 &  1999.7801 & \phn$-$56.90 & \phn$+$74.95 &   $-$5.01 &  $+$1.50 &  \phn$+$9.06  &  $-$0.35   &    0.1015 &  0.5047 & $-$0.0144 \\
  51473.6526 &  1999.8047 & \phn$+$32.28 & \phn$-$64.30 &   $-$5.45 &  $-$1.79 &  \phn$-$2.87  &  $-$0.84   &    0.9244 &  0.5063 & $-$0.0144 \\
  51477.5968 &  1999.8155 & \phn$+$48.61 & \phn$-$77.00 &   $-$3.19 &  $+$0.85 &  \phn$+$3.29  &  $+$1.40   &    0.6051 &  0.5069 & $-$0.0144 \\
  51510.5931 &  1999.9058 & \phn$+$69.96 & $-$115.42    &   $-$4.44 &  $+$0.25 &  \phn$-$4.73  &  $-$0.04   &    0.6661 &  0.5126 & $-$0.0143 \\
  51537.4847 &  1999.9794 & \phn$-$68.89 & \phn$+$85.11 &   $-$3.12 &  $-$0.24 &  \phn$+$5.52  &  $+$1.13   &    0.1257 &  0.5173 & $-$0.0143 \\
  51558.4542 &  2000.0368 & \phn$-$38.41 & \phn$+$40.50 &   $-$3.26 &  $+$1.08 &  \phn$+$1.16  &  $+$0.87   &    0.0616 &  0.5209 & $-$0.0143 \\
  51641.8900 &  2000.2653 & \phn$+$51.05 & \phn$-$85.08 &   $-$2.42 &  $-$1.32 &  \phn$+$2.47  &  $+$1.24   &    0.6169 &  0.5353 & $-$0.0140 \\
  51713.6479 &  2000.4617 & \phn$-$89.95 & \phn$+$99.16 &   $-$3.43 &  $-$0.48 &  \phn$-$8.10  &  $-$0.17   &    0.1958 &  0.5477 & $-$0.0138 \\
  51715.6836 &  2000.4673 & \phn$-$40.85 & \phn$+$37.33 &   $-$2.04 &  $-$0.13 &  \phn$-$2.83  &  $+$1.21   &    0.0633 &  0.5481 & $-$0.0138 \\
  51777.8307 &  2000.6375 & \phn$+$14.28 & \phn$-$33.17 &   $-$3.10 &  $-$4.29 &  \phn$+$8.63  &  $-$0.20   &    0.5467 &  0.5588 & $-$0.0136 \\
  51797.8286 &  2000.6922 & \phn$-$43.62 & \phn$+$35.79 &   $-$2.25 &  $-$0.04 &  \phn$-$7.84  &  $+$0.54   &    0.0686 &  0.5623 & $-$0.0135 \\
  51842.6036 &  2000.8148 & \phn$-$77.47 & \phn$+$94.87 &   $-$1.64 &  $+$0.26 &  \phn$+$4.49  &  $+$0.90   &    0.1490 &  0.5700 & $-$0.0133 \\
  51857.6308 &  2000.8559 & \phn$+$21.79 & \phn$-$42.67 &   $-$1.56 &  $+$0.23 &  \phn$+$3.68  &  $+$0.89   &    0.5527 &  0.5726 & $-$0.0133 \\
  51861.5475 &  2000.8667 & \phn$-$93.89 & $+$112.74    &   $-$3.22 &  $-$0.41 &  \phn$+$0.79  &  $-$0.79   &    0.2218 &  0.5733 & $-$0.0132 \\
  51888.5267 &  2000.9405 & \phn$+$79.34 & $-$120.56    &   $-$3.76 &  $+$0.30 &  \phn$+$5.08  &  $-$1.48   &    0.7187 &  0.5780 & $-$0.0131 \\
  51902.4960 &  2000.9788 & \phn$+$70.32 & $-$112.81    &   $-$2.21 &  $+$0.06 &  \phn$+$0.83  &  $-$0.01   &    0.6716 &  0.5804 & $-$0.0130 \\
  51921.4546 &  2001.0307 & \phn$+$80.78 & $-$129.06    &   $-$2.76 &  $+$0.13 &  \phn$-$1.02  &  $-$0.67   &    0.7507 &  0.5837 & $-$0.0130 \\
  52071.7515 &  2001.4422 & \phn$+$78.34 & $-$115.61    &   $-$1.55 &  $+$2.05 &  \phn$+$7.26  &  $-$0.30   &    0.7983 &  0.6097 & $-$0.0121 \\
  52099.9064 &  2001.5193 & \phn$+$76.42 & $-$117.95    &   $-$1.80 &  $-$0.13 &  \phn$+$5.44  &  $-$0.71   &    0.7963 &  0.6145 & $-$0.0119 \\
  52105.5866 &  2001.5348 & \phn$-$92.25 & $+$104.57    &   $-$0.43 &  $+$1.28 &  \phn$-$6.09  &  $+$0.63   &    0.2168 &  0.6155 & $-$0.0119 \\
  52121.6122 &  2001.5787 & \phn$-$31.49 & \phn$+$36.17 &   $+$1.78 &  $+$1.19 &  \phn$+$9.35  &  $+$2.75   &    0.0460 &  0.6183 & $-$0.0118 \\
  52158.5190 &  2001.6797 & \phn$+$77.10 & $-$129.27    &   $-$0.11 &  $-$2.04 &  \phn$-$1.99  &  $+$0.65   &    0.7735 &  0.6247 & $-$0.0116 \\
  52182.6433 &  2001.7458 & \phn$-$36.58 & \phn$+$34.81 &   $-$0.92 &  $+$0.38 &  \phn$+$2.44  &  $-$0.29   &    0.0539 &  0.6288 & $-$0.0114 \\
  52194.6505 &  2001.7786 & \phn$-$83.42 & $+$109.65    &   $-$0.30 &  $+$1.53 &     $+$11.30  &  $+$0.26   &    0.1707 &  0.6309 & $-$0.0113 \\
  52220.6532 &  2001.8498 & \phn$-$97.12 & $+$112.70    &   $-$0.65 &  $-$1.43 &  \phn$-$0.30  &  $-$0.24   &    0.2515 &  0.6354 & $-$0.0111 \\
  52226.5786 &  2001.8661 & \phn$+$78.36 & $-$127.09    &   $-$0.74 &  $-$0.36 &  \phn$-$0.01  &  $-$0.36   &    0.7766 &  0.6364 & $-$0.0111 \\
  52234.5686 &  2001.8879 & \phn$-$87.85 & \phn$+$98.85 &   $-$1.62 &  $-$0.15 &  \phn$-$3.07  &  $-$1.29   &    0.1814 &  0.6378 & $-$0.0110 \\
  52240.5913 &  2001.9044 & \phn$+$77.39 & $-$128.49    &   $+$1.00 &  $-$2.51 &  \phn$+$0.29  &  $+$1.30   &    0.7479 &  0.6388 & $-$0.0110 \\
  52247.5736 &  2001.9235 & \phn$+$79.01 & $-$123.51    &   $-$0.23 &  $+$0.35 &  \phn$+$3.62  &  $+$0.03   &    0.7234 &  0.6401 & $-$0.0109 \\
  52266.4777 &  2001.9753 & \phn$+$78.62 & $-$130.69    &   $-$1.33 &  $+$0.25 &  \phn$-$3.87  &  $-$1.18   &    0.7792 &  0.6433 & $-$0.0108 \\
  52274.4698 &  2001.9972 & \phn$-$91.28 & $+$103.79    &   $-$1.42 &  $-$2.68 &  \phn$+$0.86  &  $-$1.31   &    0.1850 &  0.6447 & $-$0.0107 \\
  52287.4668 &  2002.0328 & \phn$+$78.79 & $-$123.26    &   $-$1.55 &  $+$0.20 &  \phn$+$3.98  &  $-$1.52   &    0.7236 &  0.6470 & $-$0.0106 \\
  52416.7769 &  2002.3868 & \phn$+$67.27 & $-$115.52    &   $+$0.12 &  $-$1.87 &  \phn$-$0.55  &  $-$0.58   &    0.8280 &  0.6693 & $-$0.0095 \\
  52446.6840 &  2002.4687 & \phn$+$29.45 & \phn$-$61.17 &   $-$0.44 &  $-$0.89 &  \phn$+$0.56  &  $-$1.31   &    0.5727 &  0.6745 & $-$0.0092 \\
  52458.6699 &  2002.5015 & \phn$+$72.05 & $-$117.36    &   $+$3.17 &  $+$0.92 &  \phn$+$0.58  &  $+$2.23   &    0.6804 &  0.6766 & $-$0.0091 \\
  52537.6959 &  2002.7179 & \phn$-$78.33 & \phn$+$90.48 &   $+$2.81 &  $-$0.91 &  \phn$+$4.42  &  $+$1.42   &    0.3568 &  0.6902 & $-$0.0084 \\
  52538.7321 &  2002.7207 & \phn$+$76.54 & $-$122.17    &   $-$0.43 &  $+$1.36 &  \phn$+$1.80  &  $-$1.82   &    0.7983 &  0.6904 & $-$0.0084 \\
  52566.6749 &  2002.7972 & \phn$+$74.98 & $-$123.61    &   $+$2.60 &  $-$0.81 &  \phn$+$1.36  &  $+$1.05   &    0.7060 &  0.6952 & $-$0.0081 \\
  52613.5082 &  2002.9254 & \phn$+$66.53 & $-$112.28    &   $+$1.19 &  $+$0.13 &  \phn$+$0.03  &  $-$0.63   &    0.6636 &  0.7033 & $-$0.0076 \\
  52626.4669 &  2002.9609 & \phn$-$89.56 & $+$100.34    &   $+$1.01 &  $+$0.07 &  \phn$-$2.02  &  $-$0.88   &    0.1859 &  0.7056 & $-$0.0075 \\
  52638.5178 &  2002.9939 & \phn$-$89.41 & \phn$+$94.18 &   $+$1.77 &  $-$1.39 &  \phn$-$5.90  &  $-$0.19   &    0.3213 &  0.7076 & $-$0.0074 \\
  52651.4778 &  2003.0294 & \phn$+$63.56 & $-$112.08    &   $+$1.00 &  $-$0.46 &  \phn$-$2.84  &  $-$1.03   &    0.8441 &  0.7099 & $-$0.0072 \\
  52750.8421 &  2003.3014 & \phn$-$88.59 & $+$105.04    &   $+$2.82 &  $+$1.69 &  \phn$+$2.49  &  $+$0.22   &    0.1875 &  0.7271 & $-$0.0061 \\
  52788.7918 &  2003.4053 & \phn$-$77.02 & \phn$+$92.33 &   $+$6.46 &  $+$0.03 &  \phn$+$8.21  &  $+$3.65   &    0.3596 &  0.7336 & $-$0.0057 \\
  52797.7304 &  2003.4298 & \phn$-$89.02 & \phn$+$96.14 &   $+$5.91 &  $-$3.17 &  \phn$-$0.06  &  $+$3.05   &    0.1687 &  0.7352 & $-$0.0056 \\
  52891.7370 &  2003.6872 & $-$101.07    & $+$113.83    &   $+$2.46 &  $-$4.54 &  \phn$+$3.47  &  $-$0.93   &    0.2290 &  0.7514 & $-$0.0044 \\
  52912.7271 &  2003.7446 & \phn$-$88.21 & $+$100.85    &   $+$3.76 &  $-$0.74 &  \phn$+$3.07  &  $+$0.25   &    0.1738 &  0.7551 & $-$0.0041 \\
  52943.5978 &  2003.8292 & \phn$-$86.20 & $+$104.52    &   $+$3.30 &  $+$0.56 &  \phn$+$7.89  &  $-$0.38   &    0.3292 &  0.7604 & $-$0.0037 \\
  52978.5239 &  2003.9248 & \phn$-$95.01 & $+$106.13    &   $+$2.53 &  $+$0.09 &  \phn$-$1.78  &  $-$1.35   &    0.2127 &  0.7664 & $-$0.0033 \\
  53006.4893 &  2004.0013 & \phn$-$74.36 & \phn$+$79.73 &   $+$4.33 &  $-$0.60 &  \phn$+$1.33  &  $+$0.30   &    0.1300 &  0.7713 & $-$0.0029 \\
  53030.4508 &  2004.0669 & \phn$-$83.70 & \phn$+$91.83 &   $+$4.00 &  $-$0.09 &  \phn$+$0.00  &  $-$0.16   &    0.3411 &  0.7754 & $-$0.0026 \\
  53127.8133 &  2004.3335 & \phn$+$67.91 & $-$114.52    &   $+$4.48 &  $+$1.38 &  \phn$+$0.82  &  $-$0.20   &    0.8316 &  0.7922 & $-$0.0012 \\
  53181.6702 &  2004.4810 & \phn$+$75.66 & $-$131.09    &   $+$3.90 &  $-$0.23 &  \phn$-$2.60  &  $-$1.05   &    0.7824 &  0.8015 & $-$0.0004 \\
  53187.6941 &  2004.4975 & \phn$-$81.15 & \phn$+$92.63 &   $+$6.52 &  $+$0.20 &  \phn$+$4.74  &  $+$1.54   &    0.3495 &  0.8026 & $-$0.0003 \\
  53202.6395 &  2004.5384 & \phn$+$74.82 & $-$131.70    &   $+$4.91 &  $-$1.12 &  \phn$-$3.04  &  $-$0.15   &    0.7184 &  0.8052 & $-$0.0001 \\
  53275.6923 &  2004.7384 & \phn$+$62.31 & $-$103.54    &   $+$4.43 &  $+$1.39 &  \phn$+$4.79  &  $-$0.98   &    0.8495 &  0.8178 & $+$0.0010 \\
  53304.6418 &  2004.8176 & \phn$-$93.92 & \phn$+$88.17 &   $+$5.98 &  $-$2.70 &     $-$12.75  &  $+$0.44   &    0.1862 &  0.8228 & $+$0.0015 \\
  53323.6007 &  2004.8695 & \phn$-$98.17 & $+$111.57    &   $+$5.96 &  $-$0.36 &  \phn$+$1.66  &  $+$0.34   &    0.2654 &  0.8261 & $+$0.0018 \\
  53342.5272 &  2004.9214 & \phn$-$86.33 & $+$102.44    &   $+$3.49 &  $+$0.84 &  \phn$+$7.26  &  $-$2.21   &    0.3308 &  0.8294 & $+$0.0021 \\
  53357.4967 &  2004.9623 & \phn$+$72.33 & $-$127.89    &   $+$5.58 &  $-$2.28 &  \phn$-$0.35  &  $-$0.18   &    0.7100 &  0.8319 & $+$0.0023 \\
  53372.4574 &  2005.0033 & \phn$-$54.77 & \phn$+$49.58 &   $+$4.99 &  $+$0.63 &  \phn$-$1.77  &  $-$0.83   &    0.0855 &  0.8345 & $+$0.0025 \\
  53480.8610 &  2005.3001 & \phn$-$96.41 & $+$112.01    &   $+$8.87 &  $+$0.37 &  \phn$+$4.08  &  $+$2.68   &    0.2812 &  0.8533 & $+$0.0043 \\
\enddata
\tablenotetext{a}{Computed from the ephemeris in the last column of Table~\ref{tab:allelements}.}
\tablenotetext{b}{Corrections for light travel time (see text).}
\end{deluxetable}

\clearpage

\begin{deluxetable}{cccccrcc}
\tabletypesize{\small}
\tablecolumns{8}
\tablewidth{0pt}
\tablecaption{Measured times of eclipse for \V10.\label{tab:timings}}
\tablehead{\colhead{HJD}          & \colhead{$\sigma$} & \colhead{}     & \colhead{}                      & \colhead{}                         & \colhead{}  & \colhead{O$-$C}  & \colhead{} \\
           \colhead{(2,400,000+)} & \colhead{(days)}   & \colhead{Year} & \colhead{Type\tablenotemark{a}} & \colhead{Eclipse\tablenotemark{b}} & \colhead{E} & \colhead{(days)} & \colhead{Ref}}
\startdata 
  26355.245\phn\phn   &   0.021\phn\phn   &  1931.0342 &  pg &  1 &  $-$10935.0 &  $+$0.02093 &  1  \\
  26512.460\phn\phn   &   0.021\phn\phn   &  1931.4646 &  pg &  1 &  $-$10868.0 &  $+$0.00762 &  1  \\
  26925.476\phn\phn   &   0.021\phn\phn   &  1932.5954 &  pg &  1 &  $-$10692.0 &  $+$0.00714 &  1  \\
  26958.310\phn\phn   &   0.021\phn\phn   &  1932.6853 &  pg &  1 &  $-$10678.0 &  $-$0.01237 &  1  \\
  26958.335\phn\phn   &   0.021\phn\phn   &  1932.6854 &  pg &  1 &  $-$10678.0 &  $+$0.01263 &  1  \\
  27183.594\phn\phn   &   0.021\phn\phn   &  1933.3021 &  pg &  1 &  $-$10582.0 &  $-$0.00930 &  1  \\
  28692.483\phn\phn   &   0.021\phn\phn   &  1937.4332 &  pg &  1 &   $-$9939.0 &  $-$0.02197 &  1  \\
  28753.487\phn\phn   &   0.021\phn\phn   &  1937.6002 &  pg &  1 &   $-$9913.0 &  $-$0.03072 &  1  \\
  36788.465\phn\phn   &   0.021\phn\phn   &  1959.5988 &  pg &  1 &   $-$6489.0 &  $+$0.02863 &  1  \\
  36868.261\phn\phn   &   0.021\phn\phn   &  1959.8173 &  pg &  1 &   $-$6455.0 &  $+$0.03879 &  1  \\
  44166.347\phn\phn   &   0.011\phn\phn   &  1979.7983 &  v  &  1 &   $-$3345.0 &  $+$0.00712 &  2  \\
  45229.383\phn\phn   &   0.011\phn\phn   &  1982.7088 &  v  &  1 &   $-$2892.0 &  $+$0.00127 &  2  \\
  45229.385\phn\phn   &   0.021\phn\phn   &  1982.7088 &  pg &  1 &   $-$2892.0 &  $+$0.00327 &  2  \\
  45933.377\phn\phn   &   0.011\phn\phn   &  1984.6362 &  v  &  1 &   $-$2592.0 &  $-$0.00071 &  2  \\
  46001.438\phn\phn   &   0.011\phn\phn   &  1984.8226 &  v  &  1 &   $-$2563.0 &  $+$0.00756 &  2  \\
  46651.4515\phn      &   0.0015\phn      &  1986.6022 &  pe &  1 &   $-$2286.0 &  $+$0.00191 &  3  \\
  46705.420\phn\phn   &   0.011\phn\phn   &  1986.7500 &  v  &  1 &   $-$2263.0 &  $-$0.00215 &  2  \\
  47355.442\phn\phn   &   0.011\phn\phn   &  1988.5296 &  v  &  1 &   $-$1986.0 &  $+$0.00431 &  2  \\
  47362.476\phn\phn   &   0.0015\phn      &  1988.5489 &  pe &  1 &   $-$1983.0 &  $-$0.00095 &  2  \\
  48972.2716\phn      &   0.0007\phn      &  1992.9563 &  pe &  1 &   $-$1297.0 &  $-$0.00127 &  2  \\
  48972.2735\phn      &   0.0005\phn      &  1992.9563 &  pe &  1 &   $-$1297.0 &  $+$0.00063 &  2  \\
  49528.413\phn\phn   &   0.011\phn\phn   &  1994.4789 &  v  &  1 &   $-$1060.0 &  $-$0.02707 &  2  \\
  49535.4799\phn      &   0.0005\phn      &  1994.4982 &  pe &  1 &   $-$1057.0 &  $-$0.00024 &  4  \\
  49535.4802\phn      &   0.0008\phn      &  1994.4982 &  pe &  1 &   $-$1057.0 &  $+$0.00006 &  4  \\
  49941.4566\phn      &   0.0003\phn      &  1995.6097 &  pe &  1 &    $-$884.0 &  $-$0.00046 &  5  \\
  49941.4572\phn      &   0.0003\phn      &  1995.6097 &  pe &  1 &    $-$884.0 &  $+$0.00014 &  5  \\
  50286.4260\phn      &   0.011\phn\phn   &  1996.5542 &  v  &  1 &    $-$737.0 &  $+$0.00719 &  2  \\
  51159.3771\phn      &   0.0022\phn      &  1998.9442 &  pe &  1 &    $-$365.0 &  $-$0.00136 &  2  \\
  51159.3787\phn      &   0.0008\phn      &  1998.9442 &  pe &  1 &    $-$365.0 &  $+$0.00024 &  2  \\
  51159.3789\phn      &   0.0009\phn      &  1998.9442 &  pe &  1 &    $-$365.0 &  $+$0.00044 &  6  \\
  51159.3791\phn      &   0.0007\phn      &  1998.9442 &  pe &  1 &    $-$365.0 &  $+$0.00064 &  2  \\
  52015.90554         &   0.00010         &  2001.2893 &  pe &  1 &       0.0   &  $+$0.00013 &  7  \\
  52095.6909\phn      &   0.0002\phn      &  2001.5077 &  pe &  1 &     $+$34.0 &  $-$0.00029 &  8  \\
  52102.73118         &   0.00013         &  2001.5270 &  pe &  1 &     $+$37.0 &  $+$0.00008 &  8  \\
  52109.77080         &   0.00012         &  2001.5463 &  pe &  1 &     $+$40.0 &  $-$0.00022 &  8  \\
  52149.6636\phn      &   0.0003\phn      &  2001.6555 &  pe &  1 &     $+$57.0 &  $-$0.00029 &  8  \\
  52448.8603\phn      &   0.0011\phn      &  2002.4746 &  pe &  2 &    $+$184.5 &  $+$0.00031 &  9  \\
  52482.8861\phn      &   0.0003\phn      &  2002.5678 &  pe &  1 &    $+$199.0 &  $-$0.00006 &  9  \\
  52589.6558\phn      &   0.0018\phn      &  2002.8601 &  pe &  2 &    $+$244.5 &  $-$0.00210 &  9  \\
  52602.56431         &   0.00010         &  2002.8955 &  pe &  1 &    $+$250.0 &  $-$0.00006 &  9  \\
  52609.60438         &   0.00010         &  2002.9147 &  pe &  1 &    $+$253.0 &  $+$0.00012 &  9  \\
  52786.7736\phn      &   0.0014\phn      &  2003.3998 &  pe &  2 &    $+$328.5 &  $-$0.00113 &  10  \\
  52813.7610\phn      &   0.0002\phn      &  2003.4737 &  pe &  1 &    $+$340.0 &  $+$0.00007 &  10  \\
  52834.8804\phn      &   0.0003\phn      &  2003.5315 &  pe &  1 &    $+$349.0 &  $-$0.00017 &  10  \\
  52867.7336\phn      &   0.0003\phn      &  2003.6214 &  pe &  1 &    $+$363.0 &  $+$0.00027 &  10  \\
  52887.6813\phn      &   0.0014\phn      &  2003.6761 &  pe &  2 &    $+$371.5 &  $+$0.00165 &  10  \\
  52907.62597         &   0.00012         &  2003.7307 &  pe &  1 &    $+$380.0 &  $+$0.00001 &  10  \\
  53145.8083\phn      &   0.0022\phn      &  2004.3828 &  pe &  2 &    $+$481.5 &  $+$0.00014 &  11  \\
  53293.6434\phn      &   0.0011\phn      &  2004.7875 &  pe &  2 &    $+$544.5 &  $-$0.00178 &  11  \\
\enddata
\tablenotetext{a}{Technique: ph = photographic; v = visual; pe = photoelectric/CCD.}
\tablenotetext{b}{1 = primary; 2 = secondary.}
\tablerefs{
1. \cite{Strohmeier:62};
2. \cite{Kreiner:00};
3. \cite{Diethelm:86};
4. \cite{Agerer:95};
5. \cite{Agerer:96};
6. \cite{Ogloza:00};
7. \cite{Lacy:01};
8. \cite{LacySD:02};
9. \cite{Lacy:02};
10. \cite{Lacy:03};
11. \cite{Lacy:04}.}
\tablecomments{Timing uncertainties ($\sigma$) have been determined or adjusted as described in the text. O$-$C residuals are from the global solution in \S\ref{sec:orbit}.}
\end{deluxetable}
	
\clearpage

\begin{deluxetable}{cc}
\tablecolumns{2}
\tablewidth{0pt}
\tablehead{\colhead{HJD} & \colhead{$\Delta V$} \\
\colhead{(2,400,000+)} & \colhead{(mag)}}
\tablecaption{Differential $V$-band photometry for \V10 from the URSA telescope.\label{tab:ursa}}
\startdata 
   52102.73110 &  $-$0.101 \\
   52428.91322 &  $-$0.091 \\
   52095.69134 &  $-$0.091 \\
   52076.91832 &  $-$0.083 \\
   52149.66397 &  $-$0.102 \\
\enddata   
\tablecomments{Table~\ref{tab:ursa} is published in its entirety in
the electronic edition of the Astrophysical Journal. A portion is shown here for
guidance regarding its form and content.}
\end{deluxetable}

\clearpage

\begin{deluxetable}{cc}
\tablecolumns{2}
\tablewidth{0pc}
\tablehead{\colhead{HJD} & \colhead{$\Delta B$} \\
\colhead{(2,400,000+)} & \colhead{(mag)}}
\tablecaption{Differential $B$-band photometry for \V10 from Gettysburg.\label{tab:gettyb}}
\startdata 
   52548.59272 &   0.896 \\
   52541.55330 &   0.907 \\
   52548.59492 &   0.892 \\
   52548.59712 &   0.885 \\
   52541.55930 &   0.900 \\
\enddata   
\tablecomments{Table~\ref{tab:gettyb} is published in its entirety in
the electronic edition of the Astrophysical Journal. A portion is shown here for
guidance regarding its form and content.}
\end{deluxetable} 

\clearpage

\begin{deluxetable}{cc}
\tablecolumns{2}
\tablewidth{0pc}
\tablehead{\colhead{HJD} & \colhead{$\Delta V$} \\
\colhead{(2,400,000+)} & \colhead{(mag)}}
\tablecaption{Differential $V$-band photometry for \V10 from Gettysburg.\label{tab:gettyv}}
\startdata 
   52541.55170 &   0.592 \\
   52548.59332 &   0.575 \\
   52541.55390 &   0.590 \\
   52548.59552 &   0.577 \\
   52548.59782 &   0.575 \\
\enddata   
\tablecomments{Table~\ref{tab:gettyv} is published in its entirety in
the electronic edition of the Astrophysical Journal. A portion is shown here for
guidance regarding its form and content.}
\end{deluxetable} 

\clearpage

\begin{deluxetable}{cc}
\tablecolumns{2}
\tablewidth{0pc}
\tablehead{\colhead{HJD} & \colhead{$\Delta R$} \\
\colhead{(2,400,000+)} & \colhead{(mag)}}
\tablecaption{Differential $R$-band photometry for \V10 from Gettysburg.\label{tab:gettyr}}
\startdata 
   52541.55210 &   0.389 \\
   52548.59372 &   0.377 \\
   52541.55430 &   0.392 \\
   52548.59592 &   0.368 \\
   52548.59822 &   0.376 \\
\enddata   
\tablecomments{Table~\ref{tab:gettyr} is published in its entirety in
the electronic edition of the Astrophysical Journal. A portion is shown here for
guidance regarding its form and content.}
\end{deluxetable} 

\clearpage

\begin{deluxetable}{cc}
\tablecolumns{2}
\tablewidth{0pc}
\tablehead{\colhead{HJD} & \colhead{$\Delta I$} \\
\colhead{(2,400,000+)} & \colhead{(mag)}}
\tablecaption{Differential $I$-band photometry for \V10 from Gettysburg.\label{tab:gettyi}}
\startdata 
   52548.59172 &   0.191 \\
   52541.55240 &   0.187 \\
   52548.59402 &   0.190 \\
   52541.55460 &   0.172 \\
   52548.59622 &   0.185 \\
\enddata   
\tablecomments{Table~\ref{tab:gettyi} is published in its entirety in
the electronic edition of the Astrophysical Journal. A portion is shown here for
guidance regarding its form and content.}
\end{deluxetable} 

\clearpage

\begin{deluxetable}{lcccc}
\tablecaption{Comparison stars for \V10.\label{tab:comparisons}}
\tablecolumns{5}
\tablewidth{0pc}
\tablehead{\colhead{Star} & \colhead{R.A. (J2000)} & \colhead{Dec. (J2000)} & 
\colhead{$B_T$} & \colhead{$V_T$}}
\startdata
\noalign{\vskip -9pt}
\sidehead{URSA}
GSC 03600-00278  &  21:07:34.082 & +52:06:51.60 &  11.210 & 10.605  \\
GSC 03600-00377  &  21:07:22.127 & +52:01:37.07 &  12.555 & 11.824  \\
GSC 03600-00423  &  21:06:55.420 & +52:02:49.14 &  11.837 & 11.456  \\
GSC 03600-00443  &  21:06:40.440 & +52:08:02.96 &  13.087 & 11.451  \\
\sidehead{Gettysburg}
GSC 03600-00425  &  21:06:49.626 & +52:00:17.08 &   \phn9.261 &  \phn9.026  \\
GSC 03600-00259  &  21:07:40.472 & +51:52:56.94 &  11.610 & 10.318  \\
\enddata

\tablecomments{Coordinates and magnitudes $B_T$ and $V_T$ are taken from
the Tycho-2 Catalog \citep{Hog:00}.}
\end{deluxetable}

\clearpage

\begin{deluxetable}{lcccccc}
\tabletypesize{\scriptsize}
\tablecaption{Light curve solutions for \V10.\label{tab:lightelements}}
\tablecolumns{7}
\tablewidth{0pc}
\tablehead{& \multicolumn{5}{c}{Gettysburg} & URSA \\ \noalign{\vskip 3pt}
\cline{2-6} \\ \noalign{\vskip -7pt}
\colhead{\hfil~~~~Parameter\tablenotemark{a}~~~~~~} & \colhead{$B$~~~~~} & \colhead{~~~~~$V$~~~~~} & 
\colhead{~~~~~$R$~~~~~} & \colhead{~~~~~$I$~~~~~} & \colhead{Mean} & \colhead{~~~~~~~$V$~~~~~}}
\startdata
$J \equiv J_{\rm Ab}/J_{\rm Aa}$\dotfill                             & 0.398  &   0.472  &   0.525   &   0.590  & \nodata &   0.436  \\
                                            & 0.005  &   0.005  &   0.006   &   0.005  & \nodata &   0.002  \\
\noalign{\vskip 6pt}			                                                                      
$r_{\rm Aa}$\dotfill                    & 0.1635 &   0.1625 &   0.1635  &   0.1648 & 0.164   &   0.1674 \\
                                            & 0.0009 &   0.0010 &   0.0010  &   0.0009 & 0.002   &   0.0009 \\
\noalign{\vskip 6pt}			                                                                      
$r_{\rm Ab}$\dotfill                    & 0.0959 &   0.0970 &   0.0971  &   0.0994 & 0.097   &   0.1009 \\
                                            & 0.0007 &   0.0008 &   0.0008  &   0.0007 & 0.002   &   0.0006 \\
\noalign{\vskip 6pt}			                                                                      
$k\equiv r_{\rm Ab}/r_{\rm Aa}$\dotfill & 0.587  &   0.597  &   0.594   &   0.603  & 0.595   &   0.603  \\
					    & 0.003  &   0.003  &   0.003   &   0.002  & 0.007   &   0.009  \\
\noalign{\vskip 6pt}			                                                                      
$i_{\rm A}$ (deg)\dotfill               & 89.1   &   88.5   &   88.5    &   88.3   & 88.6    &   87.9   \\
					    &\phn0.5 & \phn0.3  & \phn0.3   & \phn0.3  & \phn0.4 & \phn0.3  \\
\noalign{\vskip 6pt}			                                                                      
$\ell_{\rm Aa}$\dotfill                 & 0.763  &   0.737  &   0.718   &   0.699  & \nodata &   0.743  \\
					    & 0.009  &   0.010  &   0.009   &   0.008  & \nodata &   0.016  \\
\noalign{\vskip 6pt}			                                                                      
$\ell_{\rm Ab}$\dotfill                 & 0.100  &   0.119  &   0.130   &   0.145  & \nodata &   0.113  \\
					    & 0.009  &   0.010  &   0.009   &   0.008  & \nodata &   0.016  \\
\noalign{\vskip 6pt}			                                                                      
$\ell_{\rm B}$\dotfill                  & 0.137  &   0.144  &   0.152   &   0.156  & \nodata &   0.144  \\
					    & fixed  &   fixed  &   fixed   &   fixed  & \nodata &   0.018  \\
\noalign{\vskip 6pt}			                                                                      
$x_{\rm Aa}$\dotfill                    & 0.74   &   0.50   &   0.48    &   0.35   & \nodata &   0.50   \\
					    & fixed  &   fixed  &   fixed   &   fixed  & \nodata &   0.03   \\
\noalign{\vskip 6pt}			                                                                      
$x_{\rm Ab}$\dotfill		    & 0.83   &   0.60   &   0.54    &   0.42   & \nodata &   0.60   \\
					    & fixed  &   fixed  &   fixed   &   fixed  & \nodata &   0.03   \\
\noalign{\vskip 6pt}			                                                                      
$y_{\rm Aa}$\dotfill		    & 0.33   &   0.33   &   0.33    &   0.33   & \nodata &   0.33   \\
					    & fixed  &   fixed  &   fixed   &   fixed  & \nodata &   fixed  \\
\noalign{\vskip 6pt}			                                                                      
$y_{\rm Ab}$\dotfill		    & 0.41   &   0.41   &   0.41    &   0.41   & \nodata &   0.41   \\
					    & fixed  &   fixed  &   fixed   &   fixed  & \nodata &   fixed  \\
\noalign{\vskip 6pt}
$\sigma$ (mag)\dotfill                  & 0.0071 & 0.0082   &  0.0081   &  0.0071  & \nodata &  0.0098  \\
$N_{\rm obs}$\dotfill                   &  732   &   738    &   739     &   740    & \nodata &  6129    \\
\enddata
\tablenotetext{a}{The second row for each parameter indicates the
uncertainty (standard deviation), or whether the parameter was held
fixed in the fit. All solutions adopt a mass ratio $q = 0.7266$ from
Table~\ref{tab:allelements}.}
\end{deluxetable}
	
\clearpage

\begin{deluxetable}{lcc}
\tabletypesize{\scriptsize}
\tablecolumns{3}
\tablewidth{0pc}
\tablecaption{Orbital solutions for \V10.\label{tab:allelements}}
\tablehead{
\colhead{\hfil~~~~~~~~~~~~~~Parameter~~~~~~~~~~~~~~~} & \colhead{Spectroscopic only} & \colhead{Combined}}
\startdata
\sidehead{Adjusted quantities from inner orbit (Aa and Ab)} \\
\noalign{\vskip -6pt}
~~~~$P_{\rm A}$ (days)\dotfill                                &  2.3466487~$\pm$~0.0000049         &  2.34665473~$\pm$~0.00000035        \\
~~~~$\gamma$ (\kms)\dotfill                                   &  $-7.84$~$\pm$~0.25\phs            &  $-5.67$~$\pm$~0.14\phs             \\
~~~~$K_{\rm Aa}$ (\kms)\dotfill                               &  87.96~$\pm$~0.34\phn              &  87.83~$\pm$~0.20\phn               \\
~~~~$K_{\rm Ab}$ (\kms)\dotfill                               &  120.54~$\pm$~0.73\phn\phn         &  120.87~$\pm$~0.70\phn\phn          \\
~~~~$e_{\rm A}$\dotfill                                       &  0 (fixed)                         &  0 (fixed)                          \\
~~~~$\omega_{\rm Aa}$ (deg)\dotfill                           &  \nodata                           &  \nodata                            \\
~~~~$T_{\rm A}$ (HJD$-$2,400,000)\tablenotemark{a}\dotfill    & 52015.9050~$\pm$~0.0018\phm{2222}  & 52015.89295~$\pm$~0.00058\phm{2222} \\
\sidehead{Adjusted quantities from outer orbit (Aa+Ab and B)} \\						                                
\noalign{\vskip -6pt}										                                         
~~~~$P_{\rm AB}$ (days)\dotfill                               &  \nodata                           &  5786~$\pm$~76\phn\phn              \\
~~~~$K_{\rm A}$ (\kms)\dotfill                                &  \nodata                           &  5.37~$\pm$~0.26                    \\
~~~~$e_{\rm AB}$\dotfill                                      &  \nodata                           &  0.469~$\pm$~0.032                  \\
~~~~$i_{\rm AB}$ (deg)\dotfill                                &  \nodata                           &  112.0~$\pm$~4.9\phn\phn            \\
~~~~$\omega_{\rm A}$ (deg)\dotfill                            &  \nodata                           &  97.2~$\pm$~2.8\phn                 \\
~~~~$\Omega_{\rm AB}$ (deg)\dotfill                           &  \nodata                           &  27~$\pm$~44                        \\
~~~~$T_{\rm AB}$ (HJD$-$2,400,000)\tablenotemark{b}\dotfill   &  \nodata                           &  48545~$\pm$~55\phn\phn\phn         \\
\sidehead{Other adjusted quantities} \\								                                         
\noalign{\vskip -6pt}										                                         
~~~~$\Delta\alpha^*$ (mas)\dotfill                            &  \nodata                           &  $+$1.2~$\pm$~1.9\phs               \\
~~~~$\Delta\delta$ (mas)\dotfill                              &  \nodata                           &  $+$0.40~$\pm$~0.67\phs             \\
~~~~$\Delta\mu_{\alpha}^*$ (mas/yr)\dotfill                   &  \nodata                           &  $-$2.5~$\pm$~3.5\phs               \\
~~~~$\Delta\mu_{\delta}$ (mas/yr)\dotfill                     &  \nodata                           &  $-$4.0~$\pm$~1.9\phs               \\
~~~~$\Delta\pi$ (mas)\dotfill                                 &  \nodata                           &  $-$0.26~$\pm$~0.20\phs             \\

\sidehead{Derived quantities} \\						                                         
\noalign{\vskip -6pt}										                                         
~~~~$K_{\rm B}$ (\kms)\dotfill                                &  \nodata                           &  12.85~$\pm$~0.53\phn               \\
~~~~$K_{\rm O-C}$ (days)\dotfill                              &  \nodata                           &  0.01458~$\pm$~0.00048              \\
~~~~$\mu_{\alpha}^*$ (mas/yr)\dotfill                         &  \nodata                           &  $+$17.4~$\pm$~3.5\phn\phs          \\
~~~~$\mu_{\delta}$ (mas/yr)\dotfill                           &  \nodata                           &  $+$35.5~$\pm$~1.9\phn\phs          \\
~~~~$\pi$ (mas)\dotfill                                       &  \nodata                           &  5.99~$\pm$~0.20                    \\
~~~~$M_{\rm Aa}$ (M$_{\sun}$)\dotfill                         &  (1.274~$\pm$~0.018)/$\sin^3 i_{\rm A}$&  1.282~$\pm$~0.015                \\ 
~~~~$M_{\rm Ab}$ (M$_{\sun}$)\dotfill                         &  (0.9297~$\pm$~0.0096)/$\sin^3 i_{\rm A}$&  0.9315~$\pm$~0.0068              \\ 
~~~~$M_{\rm B}$ (M$_{\sun}$)\dotfill                          &  \nodata                           &  0.925~$\pm$~0.036                  \\ 
~~~~$q\equiv M_{\rm Ab}/M_{\rm Aa}$\dotfill                   &  0.7297~$\pm$~0.0053               &  0.7266~$\pm$~0.0042                \\
~~~~$M_{\rm A}$ (M$_{\sun}$)\dotfill                          &  \nodata                           &  2.213~$\pm$~0.021                  \\ 
~~~~$M_{\rm A}+M_{\rm B}$ (M$_{\sun}$)\dotfill                &  \nodata                           &  3.139~$\pm$~0.046                  \\ 
~~~~$a_{\rm A}$ ($R_{\sun}$)\dotfill                          &  (9.666~$\pm$~0.038)/$\sin i_{\rm A}$&  9.681~$\pm$~0.031                  \\
~~~~$a''_{\rm AB}$ (mas)\dotfill                              &  \nodata                           &  55.3~$\pm$~1.8\phn                 \\
~~~~$a_{\rm AB}$ (AU)\dotfill                                 &  \nodata                           &  9.235~$\pm$~0.082                  \\
~~~~$\alpha''_{\rm ph}$ (mas)\dotfill                         &  \nodata                           &  9.55~$\pm$~0.54                    \\
\enddata
\tablenotetext{a}{Time of primary eclipse.}
\tablenotetext{b}{Time of periastron passage.}
\end{deluxetable}

\clearpage

\begin{deluxetable}{lccc}
\tablecolumns{4}
\tablewidth{0pt}
\tablecaption{Physical parameters for \V10.\label{tab:physics}}
\tablehead{
\colhead{\hfil~~~~~~~~~~~Parameter~~~~~~~~~~~} & \colhead{Primary} & \colhead{Secondary} & \colhead{Tertiary\tablenotemark{a}}}
\startdata
Mass (M$_{\sun}$)\dotfill            &  1.282~$\pm$~0.015    &  0.9315~$\pm$~0.0068       &  0.925~$\pm$~0.036  \\
Radius (R$_{\sun}$)\dotfill          &  1.615~$\pm$~0.017    &  0.974~$\pm$~0.020         &  0.870~$\pm$~0.087  \\
$\log g$\dotfill                     &  4.129~$\pm$~0.011    &  4.430~$\pm$~0.018         &  4.525~$\pm$~0.088  \\
Temperature (K)\dotfill              &  6180~$\pm$~100\phn   &  5300~$\pm$~150\phn        &  5670~$\pm$~150\phn \\
$\log L/L_{\sun}$\dotfill            &  0.533~$\pm$~0.030    &  $-$0.173~$\pm$~0.052\phs  &  $-$0.153~$\pm$~0.066\phs \\
$M_V$ (mag)\dotfill                  &  3.456~$\pm$~0.074    &  5.457~$\pm$~0.079         &  5.23~$\pm$~0.16    \\
$v \sin i$ (\kms)\dotfill            &  36~$\pm$~2\phn       &  20~$\pm$~3\phn            &  2~$\pm$~3          \\
$v_{\rm sync} \sin i$ (\kms)\dotfill &  34.8~$\pm$~0.4\phn   &  21.0~$\pm$~0.4\phn        &  \nodata            \\
Distance (pc)\dotfill                &  166.9~$\pm$~5.6\phn\phn  & 166.9~$\pm$~5.6\phn\phn  & 166.9~$\pm$~5.6\phn\phn   \\
$m-M$ (mag)\dotfill                  &  6.113~$\pm$~0.073    &  6.113~$\pm$~0.073         &  6.113~$\pm$~0.073  \\
\enddata

\tablenotetext{a}{The radius, $\log g$, and luminosity of the tertiary
were inferred from $M_V$, temperature, and bolometric corrections by
\cite{Lejeune:98}.}

\end{deluxetable}

\clearpage

\begin{deluxetable}{lcccccccc}
\tabletypesize{\scriptsize}
\rotate
\tablewidth{0pt}
\tablecaption{Parameters for eclipsing systems in the mass range of \V10.\label{tab:otherstars}}
\tablehead{

\colhead{} & \colhead{Mass} & \colhead{Radius} & \colhead{$\log g$} &
\colhead{$T_{\rm eff}$} & \colhead{$M_V$} & \colhead{Period} &
\colhead{$v \sin i$~\tablenotemark{a}} & \colhead{$\log L_X$\tablenotemark{b}} \\

\colhead{~~~~~~~~~~~~Star~~~~~~~~~~~~} & \colhead{(M$_{\sun}$)} & \colhead{(R$_{\sun}$)} & \colhead{(cgs)} &
\colhead{(K)} & \colhead{(mag)} & \colhead{(days)} &
\colhead{($\kms$)} & \colhead{(erg~s$^{-1}$)}}
\startdata
V1061 Cyg Aa\dotfill &  1.282~$\pm$~0.015   & 1.615~$\pm$~0.017 & 4.129~$\pm$~0.011  & 6180~$\pm$~100    &  3.456~$\pm$~0.074 &  \phn2.3467  &  36~$\pm$~2\phn    & 30.12  \\
V1061 Cyg Ab\dotfill &  0.9315~$\pm$~0.0068 & 0.974~$\pm$~0.020 & 4.430~$\pm$~0.018  & 5300~$\pm$~150    &  5.457~$\pm$~0.079 &  \phn2.3467  &  20~$\pm$~3\phn    &   \\
V1061 Cyg B\tablenotemark{c}\dotfill  &  0.925~$\pm$~0.036   & 0.870~$\pm$~0.087 & 4.525~$\pm$~0.088  & 5670~$\pm$~150    &  5.23~$\pm$~0.16   &  \nodata     &   2~$\pm$~3        &   \\
\noalign{\vskip 5pt}
FL Lyr A\dotfill     &  1.221~$\pm$~0.016   & 1.282~$\pm$~0.028 & 4.309~$\pm$~0.020  & 6150~$\pm$~100    &  3.95~$\pm$~0.09   &  \phn2.1782  &  30~$\pm$~2\phn    & 30.19  \\
FL Lyr B\dotfill     &  0.960~$\pm$~0.012   & 0.962~$\pm$~0.028 & 4.454~$\pm$~0.026  & 5300~$\pm$~100    &  5.37~$\pm$~0.10   &  \phn2.1782  &  25~$\pm$~2\phn    &   \\
\noalign{\vskip 5pt}
RW Lac A\dotfill     &  0.928~$\pm$~0.006   & 1.188~$\pm$~0.004 & 4.257~$\pm$~0.003  & 5760~$\pm$~100    &  4.48~$\pm$~0.09   &  10.3692     &  2~$\pm$~2 &  $< 29.6$ \\
RW Lac B\dotfill     &  0.870~$\pm$~0.004   & 0.949~$\pm$~0.004 & 4.409~$\pm$~0.004  & 5560~$\pm$~150    &  5.12~$\pm$~0.13   &  10.3692     &  0~$\pm$~2 &  \\
\noalign{\vskip 5pt}
HS Aur A\dotfill     &  0.900~$\pm$~0.019   & 1.004~$\pm$~0.024 & 4.389~$\pm$~0.023  & 5350~$\pm$~75\phn &  5.23~$\pm$~0.08   &  \phn9.8154  &  \nodata       &  $< 29.2$ \\
HS Aur B\dotfill     &  0.879~$\pm$~0.017   & 0.873~$\pm$~0.024 & 4.500~$\pm$~0.025  & 5200~$\pm$~75\phn &  5.68~$\pm$~0.08   &  \phn9.8154  &  \nodata       &   \\
\noalign{\vskip 5pt}
Sun\tablenotemark{d}\dotfill          &  1.000               & 1.000             & 4.438              & 5780              &  4.83              &  \nodata     &    2           &  26.4--27.7 \\
\enddata

\tablenotetext{a}{Measured rotational velocities for the eclipsing
systems are all consistent with synchronous rotation.  Rotational
velocity measurements for HS~Aur are not available, but
\cite{Popper:86} have shown that the lines are sharp. Synchronous
velocities for the primary and secondary are 5.2~\kms\ and 4.5~\kms,
respectively.}

\tablenotetext{b}{The values listed are for each system as a whole.}
\tablenotetext{c}{Values for the radius and $\log g$ are inferred from other properties; see \S\ref{sec:tertiary}.}
\tablenotetext{d}{The range in the X-ray luminosity of the Sun represents the change during the activity cycle \citep{Peres:00}.}

\end{deluxetable}

\clearpage

\clearpage

\clearpage

\clearpage

\clearpage

\clearpage

\clearpage

\begin{figure} 
\epsscale{0.9} 
\plotone{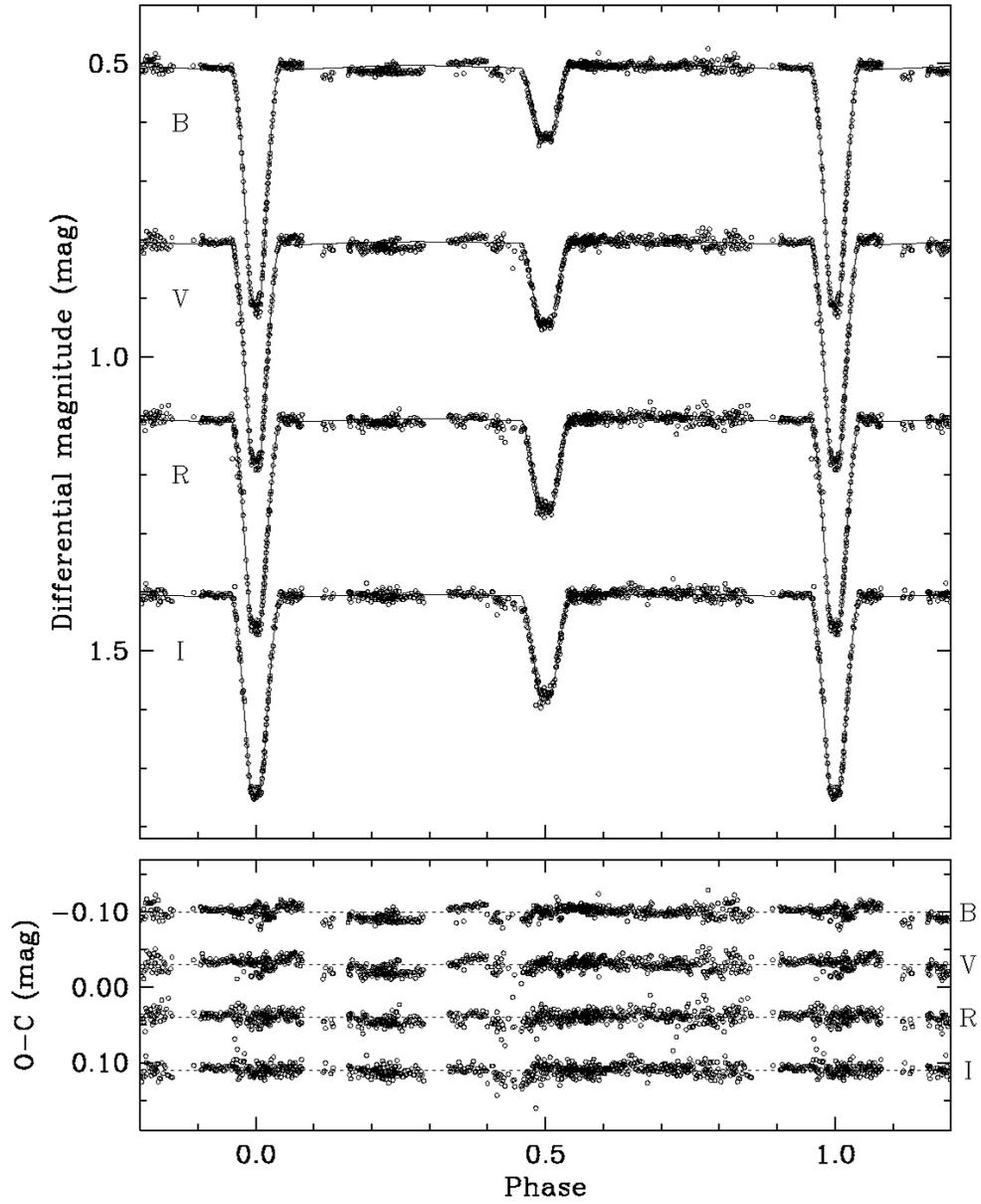}
\vskip -0.5in
 \figcaption[]{$BV\!RI$ observations for \V10 from
Gettysburg, along with our calculated light curves and residuals from
those fits. The light curves in different passbands are shifted
vertically for clarity. Patterns in the residuals near phase 0.45 may
be due to spots (see text).\label{fig:getty}}
 \end{figure}

\clearpage

\clearpage

\clearpage

\clearpage

\clearpage

\clearpage

\clearpage

\clearpage

\clearpage

\end{document}